\documentclass[prd,floatfix,showpacs]{revtex4}

\usepackage{graphicx}

\usepackage{bm}
\newcommand{\vc}[1]{\bm{#1}}

\newcommand{\wq}{{\cal Q}}

\newcommand{\Real}{{\sf R}}

\newcommand{\Complex}{{\sf C}}

\newcommand{\ieps}[1]{\includegraphics[height=3in]{#1}}

\newcommand{\Itoint}{\mbox{$\int\!\!\!\!\!-$}}

\begin{document}

\title{Time-frequency detection algorithm for gravitational wave bursts}
\author{Julien Sylvestre}
\affiliation{LIGO Project, Massachusetts Institute of Technology,\\ NW17-161, 175 Albany St., Cambridge, MA 02139, USA.\\julien@ligo.mit.edu}

\date{\today}

\begin{abstract}
An efficient algorithm is presented for the identification of short bursts of gravitational radiation in the data from broad-band interferometric detectors.
The algorithm consists of three steps: pixels of the time-frequency representation of the data that have power above a fixed threshold are first identified.
Clusters of such pixels that conform to a set of rules on their size and their proximity to other clusters are formed, and a final threshold is applied on the power integrated over all pixels in such clusters.
Formal arguments are given to support the conjecture that this algorithm is very efficient for a wide class of signals.
A precise model for the false alarm rate of this algorithm is presented, and it is shown using a number of representative numerical simulations to be accurate at the 1\% level for most values of the parameters, with maximal error around 10\%.
\end{abstract}


\maketitle

\section{Introduction} \label{intro}
A number of large laser interferometers \cite{GWProjects} are approaching sensitivities to gravitational waves in the $\sim 10-1000$ Hz frequency band that could be sufficient for the detection of astrophysical events \cite{Thorne}.
The signals from these events will be buried deep in the instrumental noise, so that unambiguous detections will be possible only with highly efficient data processing algorithms.

The focus of this article will be on transient sources of gravitational radiation, which will be defined as sources that have a relatively short duration (milliseconds to tens of seconds) and a bandwidth which overlaps at least partially with that of the interferometric detectors.
A significant number of such transient sources have been theorized, with various levels of sophistication.
For instance, the inspiral portion of coalescing compact binaries is well-understood (by post-newtonian expansion techniques \cite{inspiral}), as is the ring-down portion if a black hole results from the coalescence \cite{Leaver}, but the merger portion is understood at best only qualitatively \cite{merger,BH}.
As the mass of the binary increases, the signal-to-noise ratio of the merger portion dominates that of the inspiral and ringdown portions; the coalescence of $10 M_\odot - 1000 M_\odot$ black hole binaries could be visible at large distances, provided that the merger waveform could be detected with sufficient efficiency \cite{BH,BH2}.
The collapse of the core of massive stars could also produce detectable signals \cite{fryer_holz}; depending on the type of progenitor, bar modes, r-modes, fragmentation instabilities or black hole ringdowns could be important sources of gravitational radiation.
The details of the waveforms for most of these different mechanisms are far from known; even in the best cases, only numerical simulations covering parts of the relevant physics are available in the literature \cite{ZM}.
Hence, as it can be seen from the preceding examples, the amount of information about the gravitational wave signal from various sources varies considerably, and this variability is obviously reflected in the efficiency of the algorithms that can be designed for each class of sources.

Only minimal assumptions about the signal will be made in this paper, and therefore the principal characteristic of the algorithm to be presented will be its robustness against poor modeling of the expected signal waveform.
Stated differently, this algorithm will be moderately sensitive to a very large class of signals, by opposition to being very sensitive to only a few specific signals. 
It will correspondingly be useful to search for transient sources that do not have waveforms that are precisely predicted, and to characterize the non-Gaussian, transient component of the noise in the instruments.

It is important when designing a detection algorithm to compare its efficacy to that of other designs; the development process can stop when an algorithm that outperforms all others is discovered.
Various measures of this efficacy can be adopted, and various techniques to obtain the optimal algorithm be used, as it is discussed in section \ref{transient_detection}.
It is argued that the algorithm that is presented in this article, the {\tt TFCLUSTERS} algorithm, has a structure which is close to that of the optimal detector for a large variety of signal classes.

The {\tt TFCLUSTERS} algorithm is explained in detail in section \ref{notation}.
To summarize, it consists of the following four steps:
\begin{description} 
\item[(i)] The data $\vc{y}$ from a gravitational wave detector are transformed into a time-frequency representation with fixed time and frequency resolutions $T$ and $F$, respectively; the instantaneous power at time $t = i T$ and frequency $f = j F$, estimated from this representation, is labeled $P_{ij}(\vc{y})$.
\item[(ii)] A threshold $\eta$ is applied on the power, in order to retain only pixels with $P_{ij}(\vc{y}) > \eta$.
\item[(iii)] Clusters of pixels with power above threshold are formed by grouping pixels sharing a common side; clusters that do not conform to a fixed set of rules on their size or distance to other clusters are discarded.
\item[(iv)] A threshold on the sum of $P_{ij}(\vc{y}) - \eta$ over the surviving clusters is applied. Data segments containing clusters satisfying this threshold lead to the acceptance of the hypothesis that they contain a signal.
\end{description}

In order to understand the operating characteristics of this algorithm, a simplified version [without the clustering analysis, i.e. consisting only of steps (i), (ii) and (iv)] is shown in section \ref{optimality} to maximize  for all signals in $\Real^N$ the signal-to-noise ratio among all detectors that are based on the estimation of a lower bound of the signal power.
This simplified algorithm is especially efficient for signals with a sparse representation in the time-frequency domain.
Since most signals are expected to have pixels that present a high degree of spatial correlation in the time-frequency domain, the clustering analysis of {\tt TFCLUSTERS} is an interesting way to capitalize on that property to filter out a large portion of the noise, as it is shown in section \ref{clustering}.

An analytical method for computing the false alarm rate associated with {\tt TFCLUSTERS} is developed, with the details for the clustering analysis being presented in section \ref{clustering_FAR}.
Using the computer-generated enumeration of all the possible clusters of a certain size that can be formed, it is shown that large clusters are exponentially unlikely to occur when only noise is present.
The rate of occurrence of pairs of clusters separated by a certain distance is computed in an analogous manner.
One example of a complete analysis of the performances of {\tt TFCLUSTERS} for a short, narrow-band signal, including the optimization of its efficiency at fixed false alarm rate, is presented in section \ref{clustering_POD}.
The efficiency of {\tt TFCLUSTERS} is compared to that of the (unrealistic) ideal power detector (which assumes a knowledge of the signal duration and central frequency) as a function of the signal-to-noise ratio of the signal; at fixed probability of detection, the reduction in signal-to-noise ratio for {\tt TFCLUSTERS} is consistently less than about 30\%.

Numerical simulations (section \ref{simulations}) confirm that the analytical model for the false alarm rate is accurate in the regime of operation relevant to {\tt TFCLUSTERS}, with errors on the predicted false alarm rates around 1\% in most situations, and with maximum errors around 10\%, due to the exclusion of higher-order terms in the theoretical modeling, or to unmodeled finite size effects.

\section{Transient Detection} \label{transient_detection}
The transient detection problem consists in choosing between two possibilities: the observed data $y(t)$, $0\leq t\leq \cal{T}$, consist of a signal term $s(t)$ and a noise term $n(t)$, where the noise is additive and is assumed to be white Gaussian with zero mean and unit variance \cite{whitening}:
\begin{equation}
y(t) = s(t) + \epsilon n(t), \label{eq:noise_model_tdom}
\end{equation}
or the data are noise alone:
\begin{equation}
y(t) = \epsilon n(t). \label{eq:just_noise}
\end{equation}

In the simplest case, $s(t)$ is fixed and known.
Perhaps the most natural optimality criterion is then to maximize the probability of detection when the signal is present for the probability of detection when no signal is present being smaller than some preassigned false alarm probability.
This so-called Neyman-Pearson criterion leads to the use of the amplitude $\hat{Q}$ of the signal, optimally estimated \cite{Fisher_estimation} from the scalar product
\begin{equation}
\hat{Q} = \langle y(t), s(t) \rangle, \label{eq:match_filter}
\end{equation} 
as the statistic on which the decision between Eq. (\ref{eq:noise_model_tdom}) and Eq. (\ref{eq:just_noise}) is made.

In the case where $s(t) \in W$ for some function space $W$, one can integrate the signal distribution to reduce the problem to the simple case of discerning between two fixed probability distributions, as is done when $s(t)$ is fixed and known.
This requires the knowledge of the prior distribution $p[s(t)]$, and results in using the likelihood ratio
\begin{equation}
\Lambda[y(t)] = \frac{\int_{s(t) \in W} p[y(t)|s(t)]p[s(t)]ds(t)}{p[y(t)|0]} \label{eq:int_likeli}
\end{equation}
as the detection statistic.
When the prior $p[s(t)]$ is unknown, the choice of an optimality criterion is not as simple as for the fixed signal case; the {\it modified} Neyman-Pearson criterion maximizes the minimum of the probability of detection over all $s(t) \in W$, for the false alarm probability being below some preassigned value, and is therefore an interesting ``conservative'' choice.
It is well-known that the optimal algorithm in terms of the modified Neyman-Pearson criterion is obtained by choosing the prior $p[s(t)]$ that is least-favorable, i.e. the prior that minimizes the minimum over $W$ of the probability of detection when the signal is present, at fixed false alarm rate \cite{LeastFavorable}.
In particular, using the generalized likelihood ratio
\begin{equation}
\Lambda'[y(t)] = \max_{s(t) \in W}\frac{p[y(t)|s(t)]}{p[y(t)|0]} \label{eq:gen_likeli}
\end{equation}
as the detection statistic does not guarantee optimality \cite{GLRT}.

For signals with excellent theoretical models, such as binary neutron star or black hole inspirals, the function space $W$ is compact enough that the least-favorable prior is approximately uniform in $s(t)$, so that the generalized likelihood ratio [Eq. (\ref{eq:gen_likeli})] derives from Eq. (\ref{eq:int_likeli}) as the optimal detection statistic.
This leads to a convenient implementation \cite{Owen} which simply thresholds on the maximum of the correlation [Eq. (\ref{eq:match_filter})] over a filter bank, which is a discretely sampled version of the function space $W$.

For the case where the signal space is not simple enough to allow matched filtering, that is when the signal is not well-modeled, a number of incoherent methods have been proposed in the literature.
One of them is the so-called excess power detector \cite{epower}, which basically threshold on the power integrated over a large number of different shapes at different positions in the time-frequency plane.
In their discussion of the optimality of the excess power detector, the authors of \cite{epower} use a prior uniform in the ``whitened'' signal subspace of all waveforms of finite duration and bandwidth, which is imposed from physical arguments, and their result is therefore only a proof of optimality with respect to that particular prior. 
In particular, this prior is not necessarily least-favorable, and their result is therefore not a proof of optimality in the modified Neyman-Pearson sense.
Similarly, the author of \cite{vicere} chooses a different prior, which is uniform in the ``unwhitened'' signal space, to derive a detector which is optimal with respect to this prior, and which is similar to the excess power detector, but is perhaps better adapted to colored instrumental noise.

In addition to these detectors, a number of {\it ad hoc} methods have been proposed to solve the transient detection problem for unmodeled transients.
They are based on the analysis of patterns in the time-frequency plane \cite{tfplane,tfplane1}, or on the time-domain analysis of the data with various filters \cite{tdomain}.
Only the authors of \cite{tdomain} discuss optimality, from a numerical point of view, i.e. by using a small number of simulated signals from \cite{ZM} that are injected into noise in numerical simulations in order to compare the performances of a few different detection algorithms.

When the noise is Gaussian, the likelihood ratio in Eq.(\ref{eq:int_likeli}) can be rewritten as \cite{kailath}:
\begin{equation}
\Lambda[y(t)] = \exp\left[\Itoint_0^{\cal{T}}{y(t) \hat{s}(t) dt} - \frac{1}{2}\int_0^{\cal{T}} \hat{s}^2(t) dt \right], \label{eq:est_likeli}
\end{equation}
where $\hat{s}(t)$ is the causal minimum mean-square error estimator of $s(t)$, i.e.,
\begin{equation}
 \hat{s}(t) = E[s(t) | y(\tau)], \label{eq:baye_est}
\end{equation}
for $E[\cdot | y(\tau)]$ the expectation over the noise given the observation $y(t)$ for $\tau<t$, assuming the model of Eq.(\ref{eq:noise_model_tdom}).
The symbol $\Itoint$ represents the It\^o stochastic integral \cite{Ito}.
Of course, the computation of $\hat{s}(t)$ in Eq.(\ref{eq:baye_est}) requires the knowledge of the prior $p[s(t)]$, but the structure of Eq.(\ref{eq:est_likeli}) suggests that an efficient approach to problems for which the integration in Eq.(\ref{eq:int_likeli}) cannot be carried out effectively (due to a complex or unknown prior, for instance) might be to develop efficient estimators of the signal, and to use them as filters to test for the presence of signals in the data (i.e., to use an {\it estimator-correlator} design).

It is shown in \cite{Donoho} that (i) transforming the (discretely sampled) observations $\{y_i : i=1,...,N\}$ to a wavelet basis, (ii) applying on the transformed observations $\{\breve{y}_i\}$ the non-linearity
\begin{equation}
\check{s}_i  = \left\{ 
\begin{array}{ll}
    0 & \mbox{if $|\breve{y}_i| < \eta$} \\
    \breve{y}_i - \eta & \mbox{if $\breve{y}_i > \eta$} \\
    \breve{y}_i + \eta & \mbox{if $\breve{y}_i < \eta$},
\end{array}
\right.
\end{equation}
for a threshold $\eta \sim \epsilon\sqrt{2\log(N)/N}$, and (iii) transforming the truncated observations $\{\check{s}_i\}$ back to the time domain lead to an estimator $\{\hat{s}_i\}$ that is optimal in the sense of giving the smallest value of the maximum of the expected least-square error over a wide class of signals.
Using this estimator and preserving the structure of Eq.(\ref{eq:est_likeli}), a simple quantity to threshold on is 
\begin{equation}
\sum_{i=1}^N \left(y_i \hat{s}_i - \frac{\hat{s}_i^2}{2}\right),
\end{equation}
which can be rewritten as
\begin{equation}
\sum_{i=1}^N \left(\frac{\check{s}_i^2}{2} + \eta |\check{s}_i|\right).
\end{equation}
The algorithmic structure of {\tt TFCLUSTERS} corresponds to this model; the two principal differences are that a short-time Fourier basis is used instead of a wavelet basis, and that an additional step involving the analysis of the correlations between the non-zero components of the estimator $\check{s}$ is introduced.

\section{Algorithm} \label{notation}
It will be assumed that $N$ data corresponding to the gravitational wave strain are read and stored as real numbers into a vector $\vc{y}$.
The data are time ordered and uniformly sampled with sampling frequency $f_s$, so the $i^{\rm th}$ component $y_i$ of the vector $\vc{y}$ is the value of the measured strain at time $i/f_s$: $y_i = y(i/f_s)$.

Step (i) of {\tt TFCLUSTERS} is the construction of the time-frequency representation of the data from the spectrogram:
\begin{equation}
P_{ij}(\vc{y}) = |\tilde{y}_{ij}|^2, \label{eq:power_gauss}
\end{equation}
where $\tilde{y}_{ij}$ is the $j^{\rm th}$ component of the discrete Fourier transform of the $i^{\rm th}$ segment of data of length $M$, $\vc{y}_i = \{y_k : k=1+(i-1)M, ..., i M \}$.
$M$ is assumed to be even and to be a factor of $N$, so that $i=1, ..., N/M$, and $j=1, ..., M/2+1$.
$j=1$ and $j=M/2+1$ correspond to DC and to the Nyquist frequencies, and will not be considered anywhere below.
Consequently, the maximum number of useful pixels in the time-frequency representation will be $N_s = (M/2-1)(N/M)$.
The time resolution $T$ of the time-frequency representation is fixed and is given by $T = M / f_s$. 
The frequency resolution $F$ is simply $F = 1/T$.
In this time-frequency representation, the two hypotheses expressed in Eqs. (\ref{eq:noise_model_tdom}) and (\ref{eq:just_noise}) become $P_{ij}(\vc{y}) = |\tilde{s}_{ij} + \epsilon\tilde{n}_{ij}|^2$ and $P_{ij}(\vc{y}) = |\epsilon\tilde{n}_{ij}|^2$, respectively.

If the noise $\vc{n}$ is Gaussian and white, the power in different pixels is statistically independent.
When no signal is present, $P_{ij}$ is the sum of the square of two independent Gaussian variables with zero mean and equal variance, and is therefore exponentially distributed:
\begin{equation}
p_{P_{ij}}(P) = \frac{e^{-P/\epsilon^2}}{\epsilon^2}.
\end{equation}
When a signal is present, the Gaussian variables have non-zero mean, and the probability density function (hereafter, pdf) of the power is \cite{spectra}:
\begin{equation}
p_{P_{ij}}(P | |\tilde{s}_{ij}|^2) = \frac{e^{-(P+|\tilde{s}_{ij}|^2)/\epsilon^2} I_0\left(2\sqrt{P |\tilde{s}_{ij}|^2} / \epsilon^2\right)}{\epsilon^2}, \label{eq:rice_prob_signal}
\end{equation}
where $I_0$ is the modified Bessel function of order zero.

The spectrogram representation is of course not the only time-frequency representation available, and may not be optimal for most signal.
It is however the simplest one to work with in this exploratory work.
Wavelet bases share the independence property of the noise in the different pixels, as do any orthonormal basis, and they are known to offer better localization properties than Fourier transforms for many signals \cite{WavDonoho}.
However, their dyadic representation makes their analysis more complicated, due to the varying shape of the pixels with scale.
Another classical way to improve the spectrogram representation is to use windows and to overlap the segments.
This is a good way to reduce artifacts from Fourier transforms (e.g. edge effects) and to increase the time resolution of the time-frequency representation, but at the cost of destroying the statistical independence of the pixels.
This independence is essential for the calculations involved in the clustering analysis presented in section \ref{clustering}, and for the correct interpretation of some of the {\tt TFCLUSTERS} thresholds.
Any practical implementation of {\tt TFCLUSTERS} can, however, deal equally well with spectrograms built with windows and overlapping segments as it does with the simpler type of spectrogram described above.

\begin{figure}
\begin{center}
\ieps{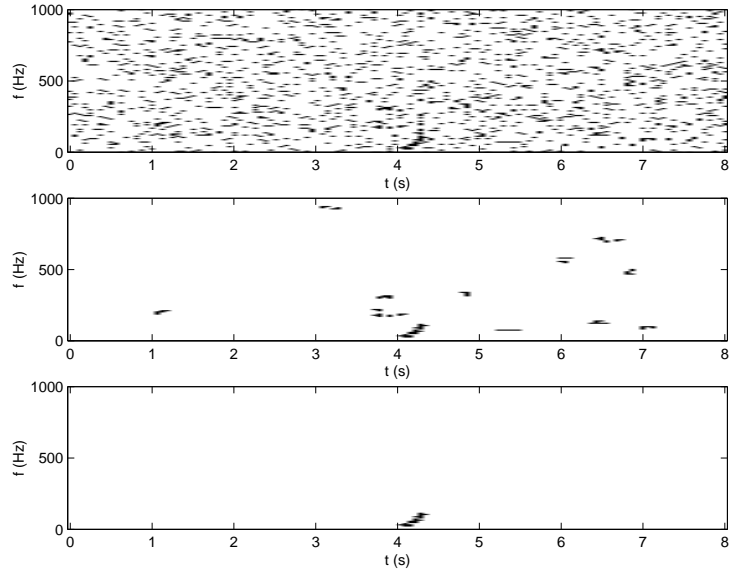}
\caption{
Examples of the various steps of the {\tt TFCLUSTERS} algorithm. From top to bottom, the time-frequency plane after Steps (ii), (iii) and (iv) is shown, for the same segment of simulated data. These data are white Gaussian noise sampled at 16384 Hz with standard deviation $5\cdot 10^{-23} {\rm Hz}^{-1/2} \sqrt{1000 {\rm Hz}}$, with a mock signal from the coalescence of a binary consisting of two 10 M$_\odot$ black holes, as described in \cite[Appendix A]{tfplane}, injected at $t = 8$ s. The binary is placed at 20 Mpc, so the signal is strong, with maximum strain amplitude $h \sim 1.7\cdot 10^{-21}$. For this example, the {\tt TFCLUSTERS} parameters are $T= 1/8$ s, $p = 0.1$, $\sigma = 5$, $\vc{\delta} = [0,0,0,0,0,0,2,3,4,4]$, and $\alpha = 0.99$.}
\label{fig:Steps}
\end{center}
\end{figure}

Step (ii) of {\tt TFCLUSTERS} consists in applying a threshold on the power $P_{ij}$.
Pixels with $P_{ij} > \eta$ are called {\it black} pixels, and other pixels are called {\it white} pixels.
The probability for any given pixel to be black when no signal is present (the {\it black pixel probability}, $p$), is given by:
\begin{equation}
p = \exp(-\eta/\epsilon^2).
\end{equation}
Figure \ref{fig:Steps} illustrates the result from Step (ii) and the effects of Step (iii) and (iv) below on simulated data.

Step (iii) of {\tt TFCLUSTERS} considers the clustering of the black pixels. 
A cluster is defined as a set of black pixels containing all the black pixels that are the nearest neighbour of any pixel in the set.
For a given pixel, its nearest neighbours are the pixels immediately to its left and to its right (time steps immediately before and after), and above and below it (frequency difference equal to the spectrogram resolution $F$).
Two pixels touching only by a ``corner'' are called {\it next} nearest neighbours.
The size $S$ of a cluster is simply defined as the number of black pixels it contains.
The notion of distance $d_c$ between two clusters $\Gamma_1$ and $\Gamma_2$ is defined as the minimal distance $d_p$ between any two pixels $p_1$ and $p_2$ in the two clusters, 
\begin{equation}
d_p(p_1, p_2) = |i_1 - i_2| + |j_1 - j_2|, \label{eq:distance}
\end{equation}
\begin{equation}
d_c(\Gamma_1, \Gamma_2) = \min_{p_1 \in \Gamma_1, p_2 \in \Gamma_2}d_p(p_1,p_2),
\end{equation}
where $(i_k,j_k)$ are the coordinates of $p_k$, i.e., $p_1$ corresponds to $P_{i_1j_1}(\vc{y})$, etc.
Hence, nearest neighbour pixels have $d_p = 1$, next-nearest neighbour pixels have $d_p=2$, and any two clusters must have $d_c \geq 2$.
This choice of distance is made for convenience, but it has the implication of making the distance isotropic in the spectrogram representation of the time-frequency domain, irrespectively of the actual values of its time and frequency resolutions.
Building a spectrogram with very long time bins and therefore very narrow frequency bins would have the effect of making the distances ``longer'' in the frequency direction than a spectrogram with short time bins and wide frequency bins.

Thresholds are applied both on the size of the cluster and on its distance to other clusters.
The latter is easily motivated for physical signals; for example, although clusters of size two that are produced by fluctuations in the noise could be likely in a certain experiment, and therefore be below the size threshold, to have two or more such clusters close from each other, say being next nearest neighbours, could be rather unlikely.
Hence, a signal with a well defined frequency that is slowly varying, so that its time-frequency track is a thin curve, could easily produce an archipelago of clusters of size two that is statistically unlikely to be produced by noise fluctuations alone.

If a cluster $\Gamma_1$ has size $S_1 \geq \sigma$, it is immediately accepted as significant.
Otherwise, its distance to all other clusters $\Gamma_2$ with size $S_2 < \sigma$ is compared to a threshold $\delta_{S_1,S_2}$ which depends explicitly on the size of the two clusters.
All the clusters with $d_c(\Gamma_1, \Gamma_2) \leq \delta_{S_1,S_2}$ are merged into a {\it generalized} cluster, which is declared significant.
If $\Gamma_2$ is already in a separate generalized cluster, $\Gamma_1$ is added to that generalized cluster.
If no cluster $\Gamma_2$ with $S_2 < \sigma$ satisfies the distance criterion, $\Gamma_1$ is rejected.
For a given choice of the minimum cluster size $\sigma$, there are $\sigma(\sigma-1)/2$ distance thresholds $\delta_{S_1,S_2}$, which will be organized below as a vector $\vc{\delta}$:
\begin{equation}
\vc{\delta} = [ \delta_{S_1,S_2} ] = [\delta_{1,1}, \delta_{1,2}, ..., \delta_{1,\sigma-1}, \delta_{2,2}, ..., \delta_{2,\sigma-1}, ..., \delta_{\sigma-1, \sigma-1}].
\end{equation}

Finally, step (iv) of {\tt TFCLUSTERS} considers significant clusters from step (iii) and threshold on their excess power $\hat{Q}$, which is defined as
\begin{equation}
\hat{Q} = \sum_{(i,j) \in \Gamma} (P_{ij}(\vc{y}) - \eta),
\end{equation}
for any given cluster $\Gamma$ of size $S$.
If no signal is present, i.e. if the cluster $\Gamma$ results from a fluctuation of the noise, the pdf of $P_{ij}(\vc{y})$ after the thresholding of step (ii) will be a truncated exponential for $(i,j) \in \Gamma$:
\begin{equation}
p_{P_{ij}(\vc{y})}(P) = \left\{ 
\begin{array}{ll}
    \frac{e^{-(P-\eta)/\epsilon^2}}{\epsilon^2} & \mbox{if $P > \eta$} \\
    0 & \mbox{otherwise.} 
\end{array}
\right.
\end{equation}
The pdf of $\hat{Q}$ will be the convolution of $S$ such distributions, i.e.
\begin{equation}
p_{\hat{Q}}(P) = \left\{ 
\begin{array}{ll}
\frac{1}{(S-1)!} \left(\frac{P-S\eta}{\epsilon^2}\right)^{S-1} \frac{e^{-(P-S\eta)/\epsilon^2}}{\epsilon^2} & \mbox{if $P > S\eta$} \\
0 & \mbox{otherwise.}
\label{eq:alpha_dist}
\end{array}
\right.
\end{equation}
Hence, setting a size-dependent threshold $\wq_S$ on $\hat{Q}$, defined by the integral equation
\begin{equation}
\alpha = \int_{S\eta}^{\wq_S} \frac{1}{(S-1)!} \left(\frac{P-S\eta}{\epsilon^2}\right)^{S-1} e^{-(P-S\eta)/\epsilon^2} \frac{dP}{\epsilon^2}, \label{eq:alpha_def}
\end{equation}
leads when only noise is present to the rejection of a fraction $\alpha$ of the clusters that survived step (iii), independently of the cluster size $S$.

Step (iv) is very similar to the scheme used in the excess power algorithm of \cite{epower}.
The essential difference is that {\tt TFCLUSTERS} chooses the pixels included in the computation of the excess power by using a threshold on the individual pixel power and a clustering analysis, while \cite{epower} use a fixed set of ``masks'' of different shapes that are translated in the time-frequency plane, and over which the power for all pixels is integrated.
The definition of these masks requires some prior expectations about the signal characteristics, and when this prior information is not sufficient to constrain the signal shapes to look for in the time-frequency plane, the selective approach of {\tt TFCLUSTERS} is expected to become more efficient than the excess power approach of \cite{epower}.

It should also be noted that other kinds of thresholds on the total cluster power could be used in step (iv).
While the one presented here penalizes in a similar fashion clusters of all sizes (it reduces the false rate by $\alpha$ independently of $S$), a size independent threshold could also be used, for instance, in which case large clusters would be more likely to survive the last step of {\tt TFCLUSTERS}.

\section{Operating Characteristics} \label{optimality}
Assessing the optimality of {\tt TFCLUSTERS} in the modified Neyman-Pearson sense is a very involved mathematical task, as it can be inferred from similar problems treated in the literature \cite{Igster}.
Nevertheless, it is shown in this section that the algorithm presents interesting properties that suggest its near-optimality for a large class of signals.
What will actually be discussed is a simplified version of the algorithm not involving the clustering analysis (step (iii) of {\tt TFCLUSTERS}).
It is assumed that the clustering analysis will improve the performances of this detector when signals that form clusters in the time-frequency plane are indeed present.

The binary test will be constructed by comparing the signal power estimate, $\hat{Q}(\vc{y}) = |\hat{\vc{s}}|^2$, to a certain threshold $\zeta$; $\hat{Q} > \zeta$ will lead to the acceptance of Eq. (\ref{eq:noise_model_tdom}).
This estimator is constructed by summing the power in the spectrogram over the pixels that have power above a certain threshold:
\begin{eqnarray}
\label{eq:power_threshold} 
|\hat{\tilde{s}}_{ij}|^2  = \left\{ 
\begin{array}{ll}
    0 & \mbox{if $P_{ij}(\vc{y}) < \eta$} \\
    P_{ij}(\vc{y}) - \eta & \mbox{otherwise} 
\end{array}
\right. \label{eq:Pthresh} \\
  \equiv (P_{ij}(\vc{y}) - \eta)_+,
\end{eqnarray}
\begin{equation}
\hat{Q}(\vc{y}) = \sum_{i,j} |\hat{\tilde{s}}_{ij}|^2. \label{eq:Qhat}
\end{equation}
The sum in Eq. (\ref{eq:Qhat}) is performed over the whole time-frequency plane, and therefore is over $N/2$ terms.

The following theorem, inspired from the work of \cite{Donoho} on signal estimation, is proved in appendix \ref{proof_th1}:\\
{\it Theorem 1:} \\
Given the model Eqs. (\ref{eq:Pthresh})-(\ref{eq:Qhat}), for $Q = |\vc{s}|^2$, and provided that $\eta = \beta\epsilon^2 \log N/2$, $\beta>1$, there exists a series of numbers $\pi_N$ with $\pi_N \rightarrow 1$ for $N \gg 1$ such that $\forall \vc{s} \in \Real^N$:\\
(i) $Pr(\hat{Q} \leq Q) = \pi_N$, \\
(ii) $Pr(\hat{Q} \geq \hat{q}) = \pi_N$, where $\hat{q}$ is any power estimator with $Pr(\hat{q} \leq Q') = \pi_N$ $\forall \vc{s}' \in \Omega(\vc{s})$, where $\Omega(\vc{s})$ is some neighbourhood of $\vc{s}$ defined in Eq. (\ref{eq:Omega}), and $Q' = |\vc{s}'|^2$.\\
(iii) $Pr\left(\hat{Q} \geq Q - \sum_{i,j}\min(2\beta\epsilon^2\log N/2, |\tilde{s}_{ij}|^2)\right) = \pi_N$.

This theorem leads to a number of observations:
\begin{itemize}
\item {\it Optimality}:
Theorem 1 shows the optimality of $\hat{Q}$ in the subset of detectors for which the condition stated in part (i) is respected: $\hat{Q}$ provides the tightest lower bound to $Q$ that can be constructed from the data, for all $\vc{s}$.

\item {\it False alarm probability}:
For a threshold $\zeta$ on $\hat{Q}$, part (i) implies that the probability to detect a signal with power $Q < \zeta$ goes to zero asymptotically with $N$ getting large.
In particular, part (i) implies that the false alarm probability goes to zero asymptotically, since when no signal is present $Q = 0$.

\item {\it False dismissal probability}:
The direct interpretation of part (iii) is that any signal with $Q > \zeta + \sum_{i,j}\min(2\beta\epsilon^2\log N/2, |\tilde{s}_{ij}|^2)$, or equivalently with $\sum_{i,j} \left(|\tilde{s}_{ij}|^2 - 2\beta\epsilon^2\log N/2\right)_+ > \zeta$, will be detected with a false dismissal probability approaching zero.
Obviously, if $\max |\tilde{s}_{ij}|^2 \leq 2\beta\epsilon^2\log N/2$, part (iii) implies that $\hat{Q} = 0$, giving a limit on how sensitive this detector can be.

\item {\it Sparse signals}:
Also as a consequence of part (iii), sparse signals that have only a small number of non-zero $\tilde{s}_{ij}$ are likely to be more easily detected, because they have a smaller value of $\sum_{i,j}\min(2\beta\epsilon^2\log N/2, |\tilde{s}_{ij}|^2)$ than signals with the same power spread over a larger number of pixels.
\end{itemize}

The scheme discussed above of thresholding on the power $P_{ij}(\vc{y})$ integrated over pixels that have power above a certain threshold $\eta$ thus possesses certain optimal properties. 
It is very efficient for signals that have a sparse representation in the time-frequency domain.
It is unlikely, however, that the optimal properties are preserved when the signal subspace is restricted to signals that form clusters in the time-frequency domain; it that case, intelligence about the spatial correlation of the signal pixels should certainly allow for more efficient algorithms.
The idea behind {\tt TFCLUSTERS} is that the {\it ad hoc} approach of merging the two-threshold scheme discussed above with a clustering analysis designed to effectively reject noise should lead to an efficient algorithm for such a signal subspace.

\section{Clustering Analysis} \label{clustering}
On physical grounds, it can be expected for most transient sources that the pixels with excess power due to the signal will tend to cluster in the time-frequency domain.
Short signals like black hole ringdowns or mergers, for instance, have durations roughly of the order of the inverse of the usable interferometer bandwidths ($\sim 1$ kHz), and therefore appear as connected clusters of duration $T$ and bandwidth equal to the search bandwidth.
For longer signals that spend tens or hundreds of cycles in the interferometers band, the required stability of the source is likely to necessitate that the dominating mechanism for the emission of the waves is governed by rotation, with the wave instantaneous frequency being around $J / \pi I$, for $J$ the magnitude of the angular momentum along the principal axis of rotation, and $I$ the moment of inertia of the source.
Complicated dynamics may lead to the spreading of the signal power over some finite frequency interval, or even the formation of sidebands well-separated from the principal frequency of the waves (in this case, each sideband will be considered an independent cluster).
When the source is not too far away, it is argued below that the amount of angular momentum radiated by gravitational waves for sources that are near the sensitivity limit of the interferometric detectors is insufficient to cause a change in the rotation frequency that is rapid enough to produce disconnected pixels in the time-frequency representation of the signal.

\begin{figure}
\begin{center}
\ieps{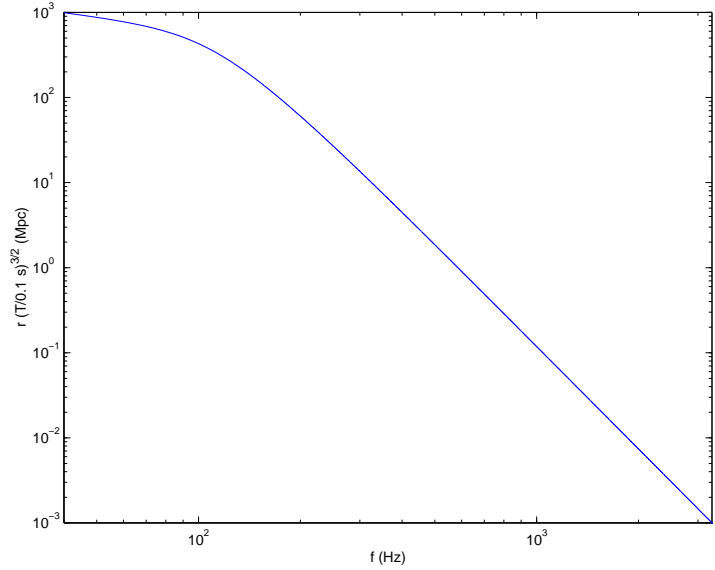}
\caption{
The maximal distance $r$ at which rotation dominated sources at the sensitivity limit of the LIGO 4k interferometers are expected to form clusters in the time-frequency plane.
}
\label{fig:range}
\end{center}
\end{figure}

Over a time interval $T$ corresponding to one time slice in the time-frequency representation, a source at a distance $r$ from the Earth that emits uniformly in all directions waves of frequency $f$ and characteristic strain amplitude $h$ will radiate a total amount of energy $\Delta E$, where \cite{Thorne}
\begin{equation}
\Delta E \sim \pi^2 \frac{c^3}{G} f^2 h^2 r^2 T.
\label{eq:deltaE}
\end{equation}
The change $\Delta J$ in angular momentum magnitude corresponding to the emission of the waves is related to the amount of radiated energy by
\begin{equation}
|\Delta J| \sim \frac{\Delta E}{\pi f} 
\label{eq:deltaJ}
\end{equation}
if most of the radiation is quadrupolar \cite{angrel}.
On the other hand, for sources with dynamics dominated by rotation, the second time derivative of the mass quadrupole moment is bounded by $\ddot{Q} \leq \pi J f$, so for $h = 2 G \ddot{Q} / r c^4$,
\begin{equation}
J \geq \frac{c^4}{2 \pi G} \frac{h r}{f}.
\label{eq:J}
\end{equation}
The value of the characteristic amplitude $h$ is expressed in term of the noise power spectral density $S_n(f)$ so that the signal-to-noise ratio for a signal with bandwidth $1/T$ is unity, corresponding to a marginally detectable source:
\begin{equation}
h = S_n^{1/2}(f) / T^{1/2}. \label{eq:hc}
\end{equation}
Combining Eqs. (\ref{eq:deltaE})-(\ref{eq:hc}):
\begin{equation}
\frac{|\Delta J|}{J} \alt 3 \cdot 10^{-4} 
\left(\frac{r}{1 {\rm Mpc}}\right)
\left(\frac{f}{100 {\rm Hz}}\right)^2
\left(\frac{S_n^{1/2}(f)}{5 \cdot 10^{-23} {\rm Hz}^{-1/2}}\right)
\left(\frac{T}{0.1 {\rm s}}\right)^{1/2},
\label{eq:dJJ}
\end{equation}
where the numbers in Eq. (\ref{eq:dJJ}) for $S_n(f)$ and $f$ correspond approximately to the minimum of the noise spectral density of the interferometers being presently developed, and the value of $T$ is chosen to match the expected time resolution of the time-frequency representations to be used on actual data.
For values of $r$ that are sufficiently small, a source does not need to radiate gravitational waves at a rate that produce variations in its angular momentum magnitude that are significant in order to be detectable.
In order of magnitude, $|\Delta J|/J \sim |\Delta f| / f$, for $\Delta f$ the change in the wave frequency over a time $T$, provided that the source doesn't change its moment of inertia by a large fraction, which is unlikely to happen over timescales of 0.1 s, except perhaps near the end of the gravitational wave signal (as e.g. in binary inspirals, at the innermost stable circular orbit), at which point the question of the amount of clustering becomes irrelevant.
Hence, the wave frequency is not expected to vary enough over the time resolution $T$ to generate pixels in contiguous time slices that are disconnected.
This can happen if
\begin{equation}
\frac{|\Delta f|}{f} \agt \frac{1}{T f}. \label{eq:disconnected_cond}
\end{equation}
Combining Eqs. (\ref{eq:dJJ}) and (\ref{eq:disconnected_cond}) gives a necessary (but not sufficient) condition for pixels to be disconnected:
\begin{equation}
r \agt 300\mbox{ Mpc} 
\left(\frac{100 \mbox{ Hz}}{f} \right)^3
\left(\frac{5 \cdot 10^{-23} {\rm Hz}^{-1/2}}{S_n^{1/2}(f)} \right)
\left(\frac{0.1 \mbox{ s}}{T} \right)^{3/2}.
\end{equation}
In this context, any rotation dominated source at the detection limit of the detector that is closer than $r$ is expected to form a cluster in the time-frequency domain.
Figure \ref{fig:range} shows the variation of $r$ with frequency, for the noise spectral density of the LIGO interferometers \cite{psdLIGO}. 
Sources with $f \alt 300$ Hz will form detectable clusters even if they are as far as the Virgo cluster; those with $f \agt 1250$ Hz, even if they are in the galaxy, may not form clusters.

These considerations show that under certain restrictions, at least two broad classes of sources (short impulsive and rotation dominated at the detection limit) should lead to signals that form clusters in the time-frequency plane.
As it will be shown below, Gaussian noise will have the opposite property, in the sense that black pixels will tend to fill the plane uniformly, without forming large clusters. 
This will be used as a powerful basis for denoising the thresholded spectrograms computed from the data.

\subsection{False Alarm Rate} \label{clustering_FAR}
The reader familiar with the mathematics of percolation theory in statistical physics has most likely recognized at this point the applicability of the results in that field to the problem under investigation here.
In particular, for a black pixel probability $p$, the average number of clusters of size $S$  per pixel of an infinite image is:
\begin{equation}
\langle n_S(p) \rangle = p^S D_S(1-p), \label{eq:ns}
\end{equation}
where $D_S$ is the so-called perimeter polynomial, and is related to the number of shapes a cluster of size $S$ can have (counting shapes related only by a translation as identical), that is to the degeneracy $g_{SR}$ of a cluster of size $S$ and perimeter $R$:
\begin{equation}
D_S(q) = \sum_{R}{g_{SR} q^R}. \label{eq:perimeter_poly}
\end{equation}
As it can be seen from Eq. (\ref{eq:perimeter_poly}), the perimeter of a cluster is the number of white pixels having at least one black pixel in the cluster as a nearest neighbour.
Coefficients of the perimeter polynomials for cluster sizes up to $S=22$ for nearest neighbours on the square lattice are tabulated in \cite{Mertens}, and were mostly generated through the use of computer enumeration techniques.
Note that $\langle n_S(p) \rangle$ is simply the probability of any pixel to be in a cluster of size $S$, divided by $S$ \cite{Conway}.

\begin{figure}
\begin{center}
\ieps{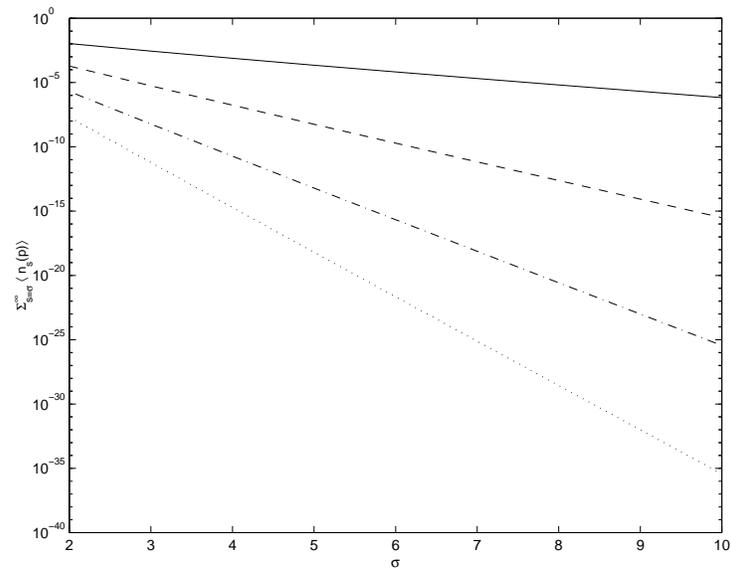}
\caption{The mean number per pixel of clusters of size greater or equal to the threshold $\sigma$ for different black pixel probabilities. Solid line: $p=0.1$, dashed: $p=10^{-2}$, dash-dotted: $p=10^{-3}$, dotted: $p=10^{-4}$.}
\label{fig:animals_size}
\end{center}
\end{figure}

For low cluster densities, the expected number of clusters per unit time $\lambda$, i.e. the cluster false alarm rate, is related for a threshold $\sigma$ on the cluster size to the time and frequency resolutions, and to the bandwidth $B$ that is searched:
\begin{equation}
\lambda =  \frac{B}{T F} \sum_{S=\sigma}^\infty \langle n_S(p) \rangle. \label{eq:rate_onecluster}
\end{equation}
In practice, because of the rapid decay of $\langle n_S(p) \rangle$ with $S$, the sum can easily be truncated without major losses of precision.
Figure \ref{fig:animals_size} illustrates the nearly exponential decay of the cluster rate with the threshold on the cluster size, for different black pixel probabilities.

In analogy to Eq. (\ref{eq:ns}), the average number per pixel of pairs of clusters of size $S_1$ and $S_2$ separated by a distance $d$ is defined as:
\begin{equation}
\langle \nu_{S_1,S_2} (d) \rangle = p^{S_1+S_2} H^d_{S_1 S_2}(1-p), \label{eq:twopoint}
\end{equation}
where the polynomial $H^d_{S_1 S_2}$ is related to the number of configurations $k_{S_1 S_2 R}(d)$ for a cluster of size $S_1$ to be within a distance $d$ from a cluster of size $S_2$, where the sum of the perimeters of the two clusters is $R$, and where configurations that are related only by a translation are again considered identical:
\begin{equation}
H^d_{S_1 S_2}(q) = \sum_R k_{S_1 S_2 R}(d) q^R.
\end{equation}
As an example, for two clusters of size 2 separated by a distance of two, Eq. (\ref{eq:twopoint}) becomes:
\begin{equation}
\langle \nu_{2,2}(2) \rangle = \frac{56 p^4 q^{11} + 40 p^4 q^{10}}{4}.
\end{equation}
The factors in the numerator account for all the possible configurations for a cluster of size 2 to be a distance 2 from another cluster of size 2, and the factor of 4 in the denominator is the total number of pixels, necessary in order not to overcount clusters related by a simple translation.
The general expression for $d>2$ is
\begin{equation}
\langle \nu_{2,2}(d) \rangle = 8 (d+1) p^4 q^{12},
\end{equation}
and hence the associated false alarm rate is
\begin{eqnarray}
\lambda_{2,2} = \frac{B}{T F} \left[ 10 p^4 q^{10} + 14 p^4 q^{11} + \sum_{d=3}^{\delta_{2,2}} 8(d+1)p^4 q^{12}  \right] \\
 = \frac{B}{T F} \left[ 10 p^4 q^{10} + 14 p^4 q^{11} + 4(\delta_{2,2}^2 + 3\delta_{2,2} - 10)p^4 q^{12} \right], \mbox{ if $\delta_{2,2} \geq 3$}, \label{eq:FAR22}
\end{eqnarray}
where $\delta_{2,2}$ is the threshold on the distance for two clusters of size 2 to be considered correlated.

\begin{figure}
\begin{center}
\includegraphics{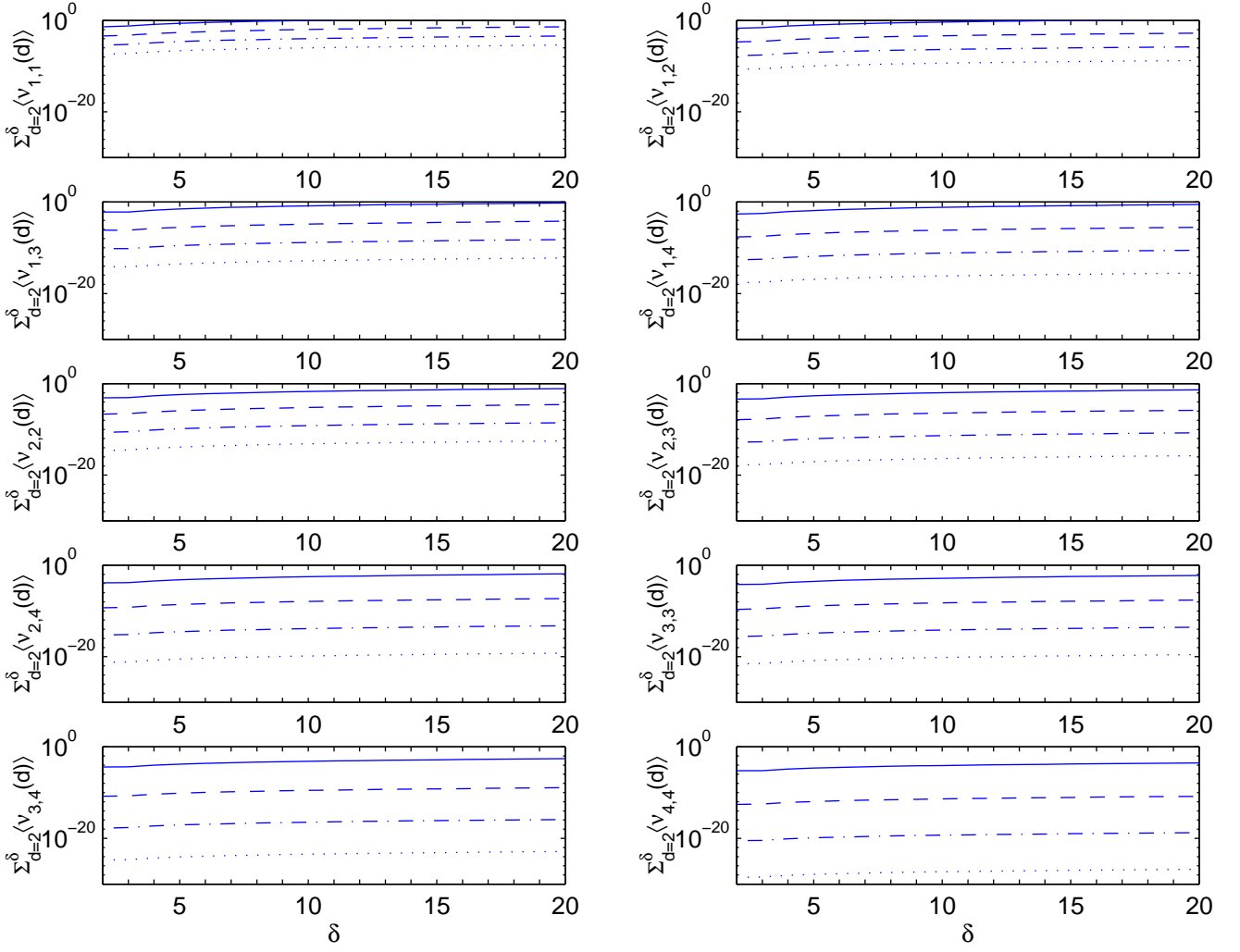}
\caption{The mean number per pixel of couples of clusters of different sizes at distance smaller or equal to the threshold $\delta$ for different black pixel probabilities. Solid line: $p=0.1$, dashed: $p=10^{-2}$, dash-dotted: $p=10^{-3}$, dotted: $p=10^{-4}$.}
\label{fig:FAR_correlations}
\end{center}
\end{figure}

Table \ref{tab:correl_coeff} gives the coefficients $k_{S_1 S_2 R}(d)$ for small cluster sizes that were obtained by computer enumeration.
General relations can easily be deduced from this table: on the square lattice, using the definition of distance from Eq. (\ref{eq:distance}), the diameter of a circle of radius $r$ is $4r$, and therefore the number of points at distance $d$ from a certain cluster increases linearly with $d$.
Moreover, when $d>2$, both clusters are guaranteed to not be sharing any perimeter white pixels, and therefore the number of configurations with fixed total perimeter consists of a ``bulk'' part (from the constant number of configurations occurring at pixels that are on the same row or column as a pixel in the cluster), and of a part growing linearly with $d$ (from the ``diagonals'' of the curve of constant distance from the first cluster).
Table \ref{tab:twopts} gives the formula for $\langle \nu_{S_1,S_2} (d) \rangle$ deduced from table \ref{tab:correl_coeff}, for $d>2$.
Values for $d=2$ can be read directly from table \ref{tab:correl_coeff}.
Figure \ref{fig:FAR_correlations} shows how the expected number of pairs of clusters of size $S_1$ and $S_2$ (i.e., $\sum_{d=2}^{\delta_{S_1,S_2}}\langle \nu_{S_1,S_2} (d) \rangle$) increases when the threshold $\delta_{S_1,S_2}$ is increased.
As expected, it also shows that pairs of large clusters are more unlikely to form than pairs of smaller clusters, and that the expected number of pairs of clusters decreases when the black pixel probability is reduced.

It should be noted that {\tt TFCLUSTERS} includes higher order terms ($n$-cluster configurations, $n \geq 3$) into its definition of generalized clusters.
These complex configurations are built by merging all the clusters satisfying the distance criteria, and
consequently the sum of Eq. (\ref{eq:rate_onecluster}) and of its equivalent for Eq. (\ref{eq:twopoint}) overestimates slightly the true false alarm rate, especially when the distance thresholds $\delta_{S_1,S_2}$ are large for the density of small clusters.

\subsection{Efficiency} \label{clustering_POD}
As shown above, the false alarm rate is computed analytically; when $\alpha = 0$ (cf. step (iv) of {\tt TFCLUSTERS}), the probability of detection of a certain signal can also be computed analytically.
It is not possible at this time to model the detection efficiency when $\alpha > 0$
because the equivalent to Eq. (\ref{eq:alpha_dist}) when a non-null signal is present is not known analytically, and is required to compute the probability of detection of a given signal.
The sensitivity and false alarm rate of {\tt TFCLUSTERS} depend on a number of parameters: $F, T, p, \sigma$ and $\vc{\delta}$.
Using some general information about the expected waveform, it is possible to run an optimization over all these parameters to maximize the probability of detection at fixed false alarm rate.
More generally, it is possible to run multiple versions of the detector, each one for a different set of parameters, in order to cover as many different classes of signal as possible.
The number of classes that can be considered is limited by the computational power available for the analysis, and is somewhat limited by the fact that results are not independent.

For comparison, the probability of detection using the ideal power detector can also be computed. 
This detector assumes that the bandwidth and duration of the signal are known and that their product is the time-frequency volume $V$, and it compares the excess power computed according to these parameters to a threshold $\kappa$.
In order to get a false alarm rate rather than a false alarm probability, it is assumed that the detector is applied with frequency $1/\tau$ on segments of independent data of length $\tau$, and that all frequency bins in the search bandwidth $B$ are considered in disjoint groups having bandwidth $\phi$.
This detector is incoherent and is therefore not as efficient as matched filtering, but it still performs better than any detector that can be implemented when neither the duration nor the bandwidth is known.
The false alarm rate is:
\begin{equation}
\lambda = \frac{B}{\phi\tau}\int_{\kappa}^\infty \frac{(P/\epsilon^2)^{V-1}e^{-P/\epsilon^2}}{\epsilon^2(V-1)!} dP,  \label{eq:false_alarm_power}
\end{equation}
and the probability of detection is \cite{spectra}
\begin{equation}
\beta = \int_{\kappa}^\infty \left( \frac{P}{V\rho^2\epsilon^2} \right)^{(V-1)/2} \exp \left(-\frac{P+V\rho^2\epsilon^2}{\epsilon^2}\right) I_{V-1}\left(2\sqrt{\frac{V\rho^2P}{\epsilon^2}}\right) \frac{dP}{\epsilon^2}. \label{eq:detection_prob_power}
\end{equation}
The case where one or more of the bandwidth, duration, central frequency or arrival time parameters have to be searched over is more complicated to analyze:  the probabilities are not statistically independent, because of the need to search over overlapping regions in the time-frequency representation of the signal, and therefore the computation of the false alarm rate is rather involved.

An example is now presented to make things more precise regarding how well {\tt TFCLUSTERS} compares to the ideal power detector.
The signal is taken to be of duration $6T$ and bandwidth smaller than $F=1/T$ (since the problem is symmetric under the interchange of the time and frequency axes, the results below are equivalent to the case of a short signal of duration $T$ and bandwidth $6/T$).
It is assumed that the signal power is distributed uniformly over its full duration and bandwidth.
As a definite example, let the bandwidth of the search be chosen to be $B=512F$, and the false alarm rate be $\lambda = 1/3600T$.
The central frequency is assumed to be such that the power is concentrated in a single constant frequency row of the spectrogram, while both the case of synchronized (the signal covers six pixels) and the case of random arrival time (the signal covers seven pixels) are considered.
For the ideal power detector, $V = 6$ or 7, and $\tau$ is set to $6T$.
For the {\tt TFCLUSTERS} algorithm, the various thresholds ($\eta$, $\sigma$ and $\delta_{S_1,S_2}$, $\alpha$ being zero) are optimized for every value of the signal-to-noise ratio in order to maximize the probability of detection (hereafter, POD) for the constraint $\lambda \leq 1/3600T$.
The details of the calculation are presented in appendix \ref{narrow_band}.
Figure \ref{fig:SNR_POD} shows a comparison of the optimized POD as a function of signal to noise ratio $\rho$, for {\tt TFCLUSTERS} and the ideal power detector.
Only two sets of clustering analysis thresholds cover the whole range of signal-to-noise ratios of figure \ref{fig:SNR_POD}, illustrating the relative independence of the performances of {\tt TFCLUSTERS} on its numerous parameters.
Given the fact that the chosen signal was a line in the time-frequency plane, and was therefore the shape that is the easiest to ``break'' by changing a black pixel into a white pixel, the example presented here can be considered as a difficult situation; most other distributions of the power (i.e. other signal shapes) would make the performances of {\tt TFCLUSTERS} and of the ideal power detectors only closer.

\section{Numerical Simulations} \label{simulations}
Extensive numerical simulations were carried out in order to confirm the validity of the analyses presented in the previous sections, and in order to explore properties of {\tt TFCLUSTERS} that are hard to study analytically.
All of the results below were produced using the same method: segments of Gaussian white noise of unit variance are produced using a random number generator \cite{RAN2}.
The segments are of duration $10^4$ s, and are sampled at 2048 Hz.
These numbers are chosen so that processing a single segment uses most of the RAM of the computers on which the code is running.
On every segment, the implementation of {\tt TFCLUSTERS} within the LIGO Scientific Collaboration Algorithm Library \cite{LAL} is used to generate a list of significant clusters, according to some pre-specified values of the various thresholds.
The 90\% central confidence interval for the rate of significant clusters in the data, assuming they form a Poisson process, is computed based on the number of detected clusters, using the standard Neyman construction \cite{Neyman}.
If the ratio of the width of the 90\% confidence interval to its central value is smaller than 1\%, the simulation terminates.
Otherwise, a number of other $10^4$ s segments are processed, until the above termination criterion is met.
For a true rate of $\lambda$, the termination criterion requires to process of order $10^5/\lambda$ worth of simulated data.

\begin{figure}
\begin{center}
\ieps{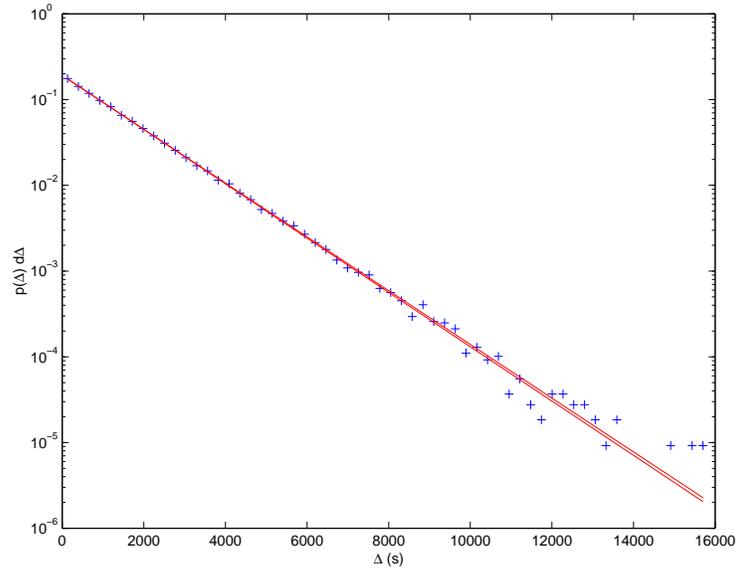}
\caption{
The probability density $p(\Delta)$ of the time delay $\Delta$ between two successive clusters, as measured empirically for $\eta = 3.719\epsilon^2$, $\sigma = 5$, $\vc{\delta} = [0,0,0,0,0,0,2,3,4,4]$, and $\alpha = 0$, for $T=1/32$ s, $B = 992$ Hz. The two continuous lines correspond to the extrema of the predicted Poisson distribution, assuming the value of the rate at both ends of its 90\% confidence interval ($(7.28 \pm 0.02)\cdot 10^{-4}$ Hz).
}
\label{fig:Poisson}
\end{center}
\end{figure}

It should be noted that the timing distribution of clusters from white noise is indeed very well approximated by a Poisson distribution.
This is confirmed in one specific case by figure \ref{fig:Poisson} which shows that the distribution of the time delay $\Delta$ between two successive clusters follows an exponential distribution, which is the defining property of a Poisson process.

The simulations were performed on a cluster of workstations with 1GHz Intel Pentium III ``Coppermine'' processors, 512 Mbytes of PC-133 RAM memory, with the Linux operating system. 
For a large range of parameter values, the processing time per CPU was on average 250 to 550 times shorter than the duration of the data segment.
Most of the time was spent at grouping the black pixels into clusters, and quite logically the most important factor in determining the speed of {\tt TFCLUSTERS} was the black pixel probability; independently of the other parameters, the ratio of processing to real time was around 300 for $\eta/\epsilon^2 = 2$, around 500 for $\eta/\epsilon^2 = 4$, and was increasing almost linearly with the power threshold $\eta$.

\subsection{The Case $\vc{\delta} = \vc{0}$, $\alpha = 0$}
\begin{figure}
\begin{center}
\ieps{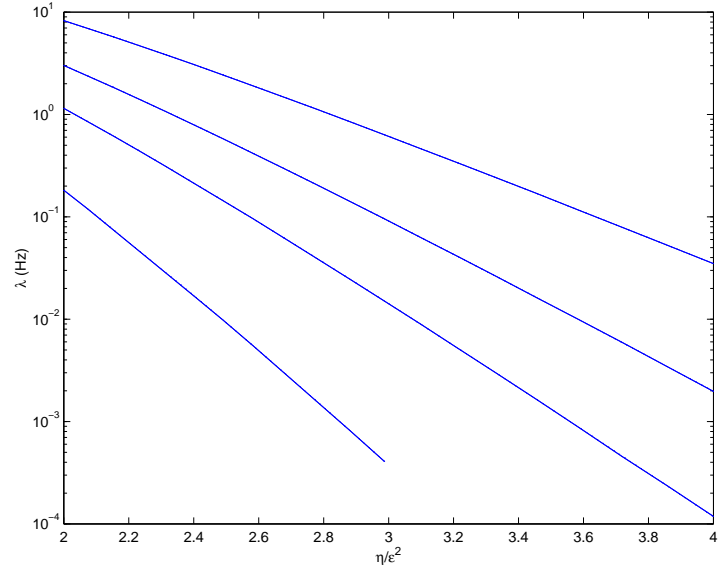}
\caption{
The measured rate $\lambda$ in white noise for various values of $\eta$, for $\vc{\delta} = \vc{0}$, $\alpha = 0$, $T=1$ s, $B=1023$ Hz.
From top to bottom, the curves correspond to $\sigma = 3,4,5,7$.
Both the error bars on $\lambda$ and the predicted rates from Eq. (\ref{eq:rate_onecluster}) are occulted by the thickness of the line.
}
\label{fig:d0}
\end{center}
\end{figure}

\begin{figure}
\begin{center}
\ieps{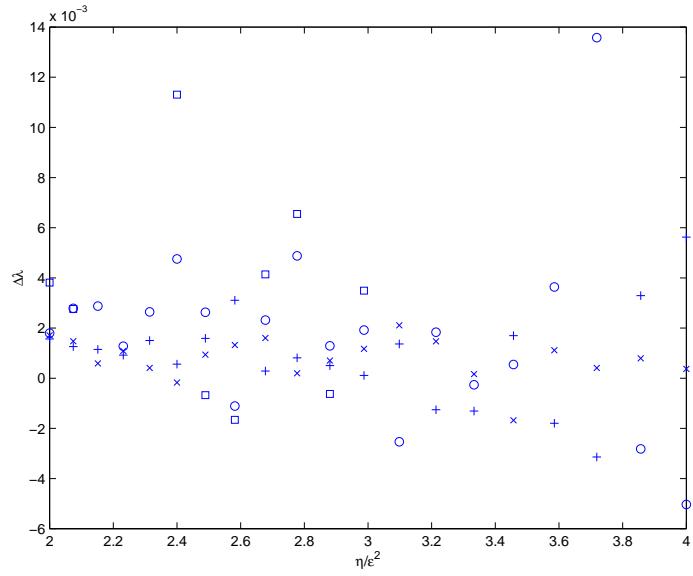}
\caption{
The fractional residuals $\Delta\lambda = (\lambda_{\rm pred} - \lambda) / \lambda_{\rm pred}$, where $\lambda_{\rm pred}$ are the predicted rates, corresponding to figure \ref{fig:d0}.
X-marks, plus signs, circles and squares correspond respectively to $\sigma = 3,4,5,7$.
}
\label{fig:d0_res}
\end{center}
\end{figure}

\begin{figure}
\begin{center}
\ieps{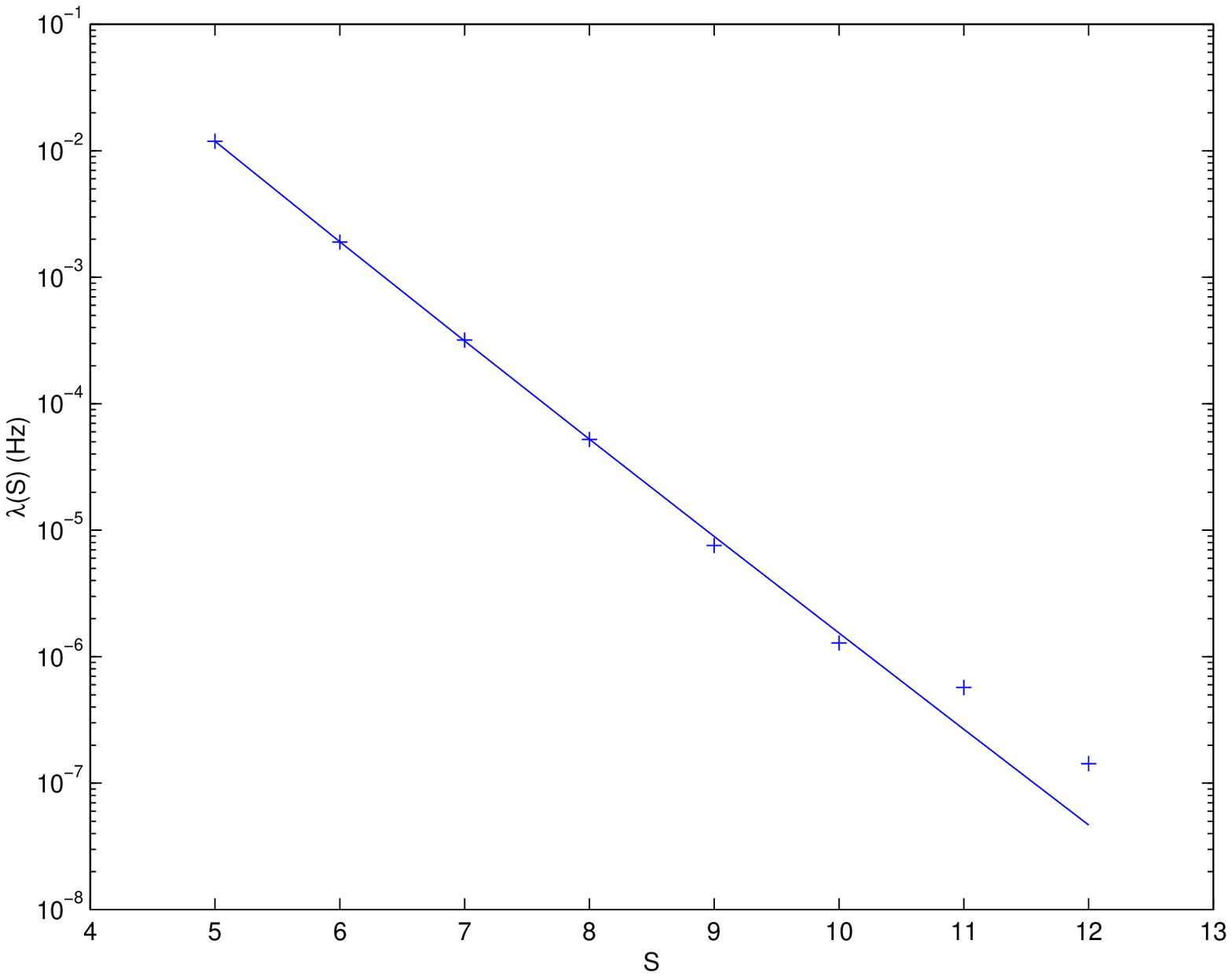}
\caption{
The measured rate $\lambda(S)$ as a function of the cluster size $S$ for $\eta = 3 \epsilon^2$, $\sigma = 5$, $\vc{\delta} = \vc{0}$, $\alpha = 0$, $T=1$ s, $B=1023$ Hz.
The plus signs are results from simulations, and the continuous line corresponds to the predictions.
This plot was constructed from $7\cdot 10^6$ s worth of simulated data.
}
\label{fig:d0_histo}
\end{center}
\end{figure}

When $\vc{\delta} = \vc{0}$ and $\alpha = 0$, the expected rate is given by Eq. (\ref{eq:rate_onecluster}).
Figures \ref{fig:d0} and \ref{fig:d0_res} show the excellent agreement between rates from simulations and predictions from Eq. (\ref{eq:rate_onecluster}).
The two agree to better than 0.5\% most of the time, commensurably with the precision of the simulations.
The sum in Eq. (\ref{eq:rate_onecluster}) is of course dominated by the first few terms, but values of $\langle n_S(p) \rangle$ as predicted from Eq. (\ref{eq:ns}) describe the simulated data also very well for large values of $S$, as shown in figure \ref{fig:d0_histo}.

\subsection{The Case $\vc{\delta} \neq \vc{0}$, $\alpha = 0$}
\begin{figure}
\begin{center}
\ieps{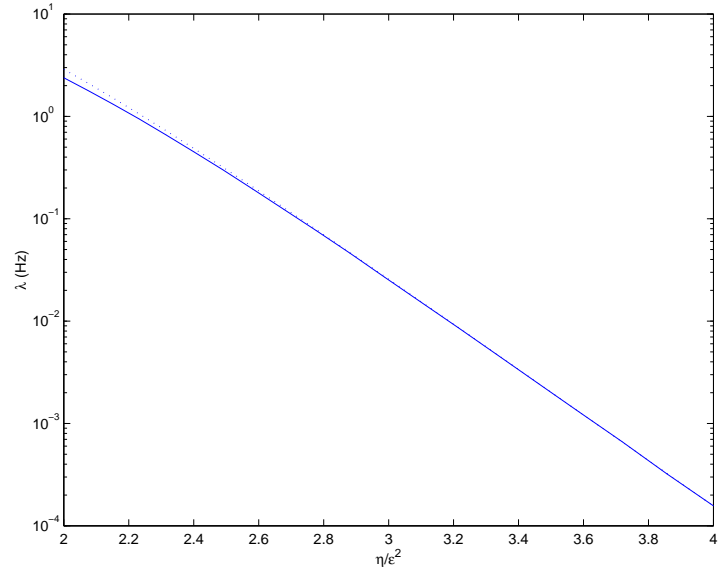}
\caption{
The measured rate $\lambda$ in white noise for various values of $\eta$, for $\sigma = 5$, $\vc{\delta} = [0,0,0,0,0,0,2,3,4,4]$, $\alpha = 0$, $T=1$ s, $B=1023$ Hz.
The dotted line correspond to the predicted rate.
}
\label{fig:dneq0}
\end{center}
\end{figure}

\begin{figure}
\begin{center}
\ieps{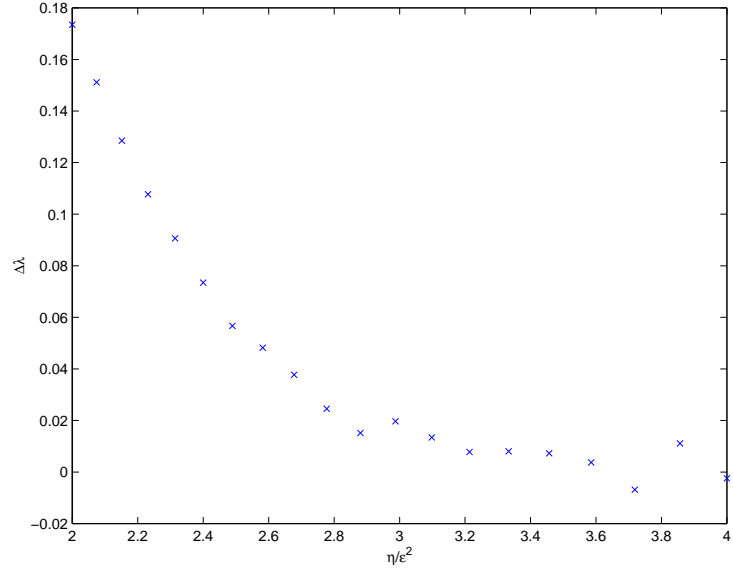}
\caption{
The fractional residuals defined as in figure \ref{fig:d0_res}, but corresponding to figure \ref{fig:dneq0}.
}
\label{fig:dneq0_res}
\end{center}
\end{figure}

For $\vc{\delta} \neq \vc{0}$, there is a contribution to the cluster rate from both Eq. (\ref{eq:rate_onecluster}) and its equivalent for $\langle\nu_{S_1,S_2}(d)\rangle$.
Figures \ref{fig:dneq0} and \ref{fig:dneq0_res} show the good agreement between rates from simulations and predictions.
When $\eta/\epsilon^2 \agt 2.8$, the agreement is at the precision level of the simulations, although there is a systematic overestimation of the measured rate by the predictions for $\eta/\epsilon^2 \alt 3.7$.
This overestimation, however, reaches almost 20\% for $\eta/\epsilon^2 = 2$; this is expected, as explained in section \ref{clustering}, because for $\eta/\epsilon^2 \alt 2.8$ the cluster density is high enough that higher-order combinations above the 2-cluster one are likely to be produced.

\begin{figure}
\begin{center}
\ieps{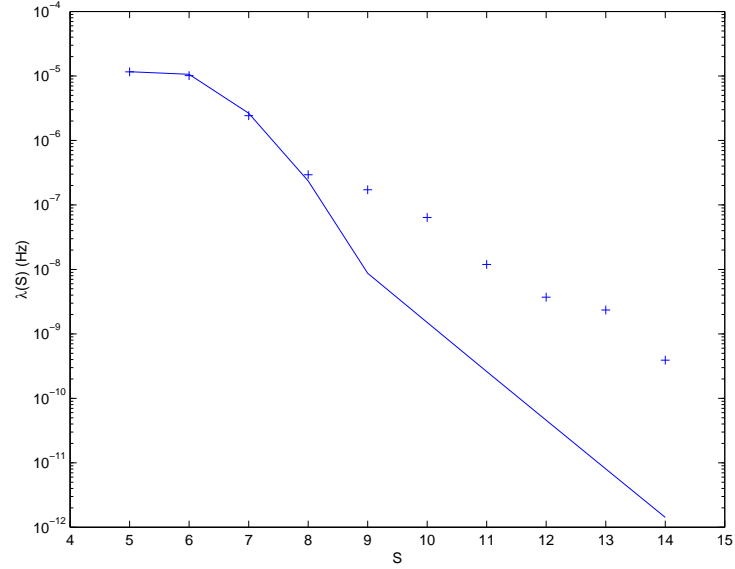}
\caption{
The measured rate $\lambda(S)$ as a function of the cluster size $S$ for $\eta = 3 \epsilon^2$, $\sigma = 5$, $\vc{\delta} = [0,0,0,0,0,0,2,3,4,4]$, $\alpha = 0$, $T=1$ s, $B=1023$ Hz.
The plus signs are results from simulations, and the continuous line corresponds to the predictions.
This plot was constructed from $5\cdot 10^6$s worth of simulated data.
}
\label{fig:dneq0_histo}
\end{center}
\end{figure}

\begin{figure}
\begin{center}
\ieps{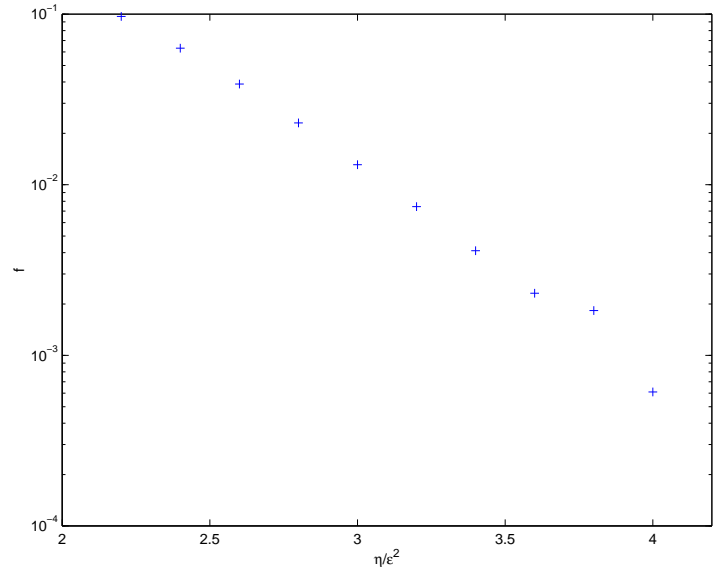}
\caption{
The ratio $f$ of the number of generalized clusters containing three or more simple clusters to the total number of clusters, as a function of the threshold $\eta$, for the same parameters as in figure \ref{fig:dneq0}.
}
\label{fig:fraction}
\end{center}
\end{figure}

To demonstrate that the error is indeed due to higher-order terms, a histogram of the contribution $\lambda(S)$ to the total rate of generalized clusters of size $S$ is presented in figure \ref{fig:dneq0_histo}, together with the theoretical prediction based on 1- and 2-cluster configurations.
The predictions systematically overestimate the measured rate for $S<8$, and underestimate it for $S\leq8$.
It should be noted that for $\vc{\delta} = [0,0,0,0,0,0,2,3,4,4]$, 3-cluster configurations can have sizes $8 \leq S \leq 12$, while 4-cluster configurations can have sizes $10 \leq S \leq 16$.
Hence, one can expect a small error for $S=8$ corresponding to the 3-cluster $(S_1,S_2,S_3) = (2,2,4)$, and no errors from higher-order terms for $S<8$, i.e. for the terms dominating the sum leading to the prediction for $\lambda$.  
Overall, more small clusters than expected get merged into generalized clusters due to these high-order terms, and the total rate is consequently overestimated.
Figure \ref{fig:fraction} quantifies the importance of high-order terms as a function of the cluster density, i.e. as a function of the threshold $\eta$; as expected, the relative importance of high-order terms becomes smaller than the 1\% level around $\eta/\epsilon^2 = 3$, in agreement with the results in figure \ref{fig:dneq0}.

\subsection{The Case $\alpha > 0$, and Finite-Size Effects}
\begin{figure}
\begin{center}
\ieps{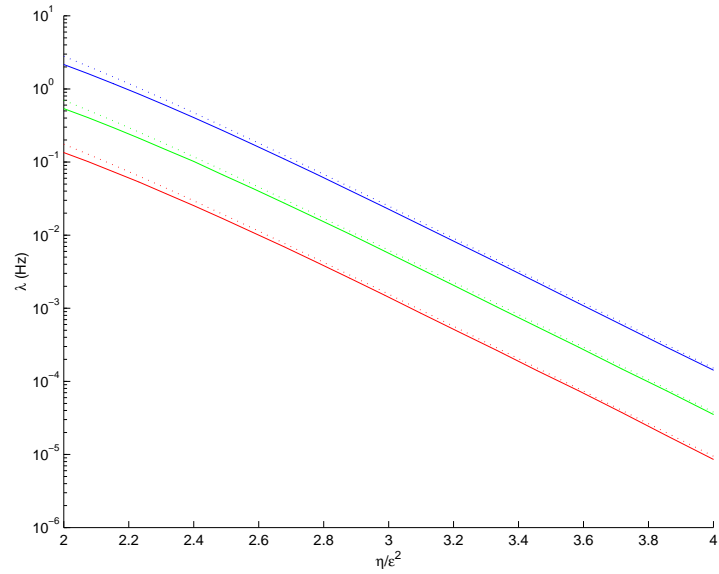}
\caption{
The measured rate $\lambda$ in white noise for various values of $\eta$, for $\sigma = 5$, $\vc{\delta} = [0,0,0,0,0,0,2,3,4,4]$, $T=1/32$ s, $B=992$ Hz.
From top to bottom, the curves correspond to $\alpha = 0, 3/4, 15/16$.
The dotted line correspond to the predicted rate.
}
\label{fig:alphaneq0}
\end{center}
\end{figure}

For $\alpha > 0$, the rate is expected to be $(1-\alpha)\lambda_0$, where $\lambda_0$ is the value of the rate for the same parameters, except that $\alpha = 0$.
Figure \ref{fig:alphaneq0} illustrates the reduction in $\lambda$ as $\alpha$ is varied.
The results are as expected from Eq. (\ref{eq:alpha_def}), within the errors from the simulations.

\begin{figure}
\begin{center}
\ieps{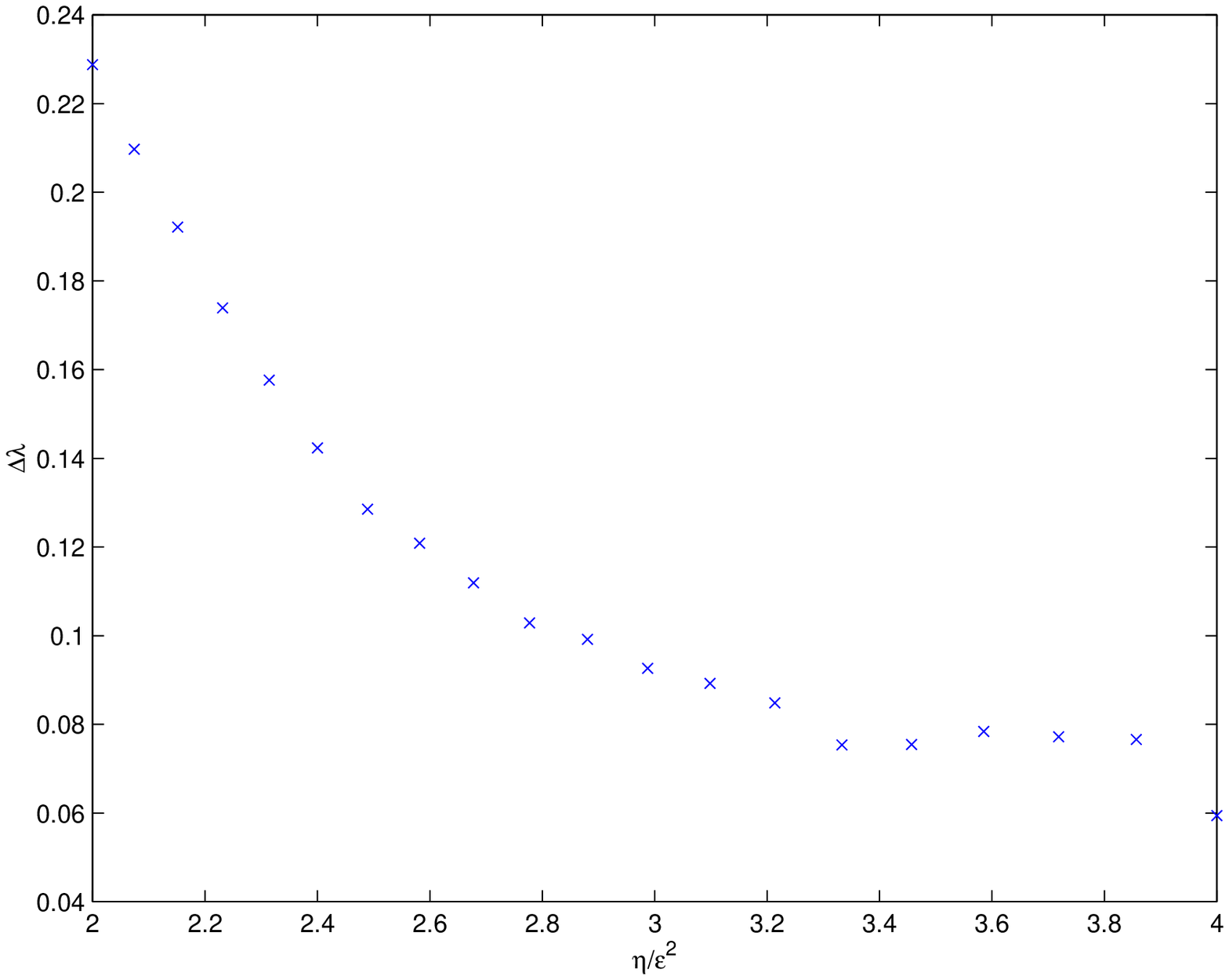}
\caption{
The fractional residuals defined as in figure \ref{fig:d0_res}, but corresponding to figure \ref{fig:alphaneq0}, for the curve with $\alpha = 0$.
}
\label{fig:alphaneq0_res}
\end{center}
\end{figure}

The simulations presented in figure \ref{fig:alphaneq0} were carried out with a very asymmetrical spectrogram: $T$ was chosen to be 1/32 s, so $F = 32$Hz, and the 992 Hz bandwidth was covered by only 31 pixels, while a segment had $32\cdot 10^4$ bins in time.
Because of that, significant finite size effects are expected; the rate predictions are based on the assumption that the time-frequency plane is infinite, but in the present case it is small enough in the frequency dimension that clusters are likely to be ``clipped'' and therefore to be smaller then expected.
Consequently, it is expected that the predictions will again overestimate the measured rate, and this is what is observed in figures \ref{fig:alphaneq0} and \ref{fig:alphaneq0_res}.
As it can be seen from figure \ref{fig:alphaneq0_res}, the fractional error from the prediction is still smaller than about 10\% when the cluster density is low enough.

\section{Conclusions}
{\tt TFCLUSTERS}, a new time-frequency detector for bursts of gravitational radiation in broad-band interferometric observatories, was described in some details in this paper.
The behavior of the detector when applied to white Gaussian noise in the absence of signals was carefully analyzed, leading to a formalism for the computation of the false alarm rate of {\tt TFCLUSTERS} for any values of its many parameters.
The results from numerical simulations showed that this analysis was accurate in most situations: errors at the 1\% level or better were obtained in ``ideal'' situations (low cluster density, large number of frequency bins), and errors of the order of 10\% appeared in situations were the analysis was expected to be less accurate, due to high order terms not included in the sums over the cluster configurations, or due to finite size effects from the limited bandwidth of the search.
In the case where the errors were large, the analysis presented in this paper was systematically overestimating the false alarm rates; should they prove from practical work to be necessary, more accurate estimates could therefore be obtained from more careful calculations, or from larger scale numerical simulations.
While the false alarm properties of {\tt TFCLUSTERS} are well-understood, the efficiency of the detector is subject to more uncertainties.

A calculation that was presented in Appendix \ref{narrow_band} showed that the efficiency of {\tt TFCLUSTERS} was comparable to that of the ideal power detector, unfortunately illustrating at the same time the mathematical complexity associated with producing such an estimate of the efficiency of the detector for a given signal.
Nevertheless, the fundamental approach used by {\tt TFCLUSTERS}, namely the use of an adaptive power integral over pixels with excessive power, was shown to be maximizing asymptotically the estimate of the power in the signal, provided it is not overestimating it, over all possible estimators of the same quantity.
This naturally suggests optimal properties for {\tt TFCLUSTERS} as a detector.
However, this can be nothing more than a conjecture at this point, as the actual proof of the optimality of {\tt TFCLUSTERS} in the (modified) Neyman-Pearson sense is of great difficulty.

Independently of the question of the optimality of {\tt TFCLUSTERS}, the structure of the algorithm is particularly practical for its implementation for the analysis of actual data.
The first power threshold on individual pixels (step (ii)) can be chosen to be frequency dependent in order to allow the analysis of colored Gaussian noise when whitening of the data is not convenient; the errors on the black pixel probability that are introduced by the non-zero correlations coloring the noise are generally negligible.
Moreover, frequency bands containing spurious interferences are easily left out of the analysis, with minor modifications to the algorithm presented in this paper.

The search for gravitational waves will of course require the operation of {\tt TFCLUSTERS} in coincidence at sites that are geographically separated and the use of information from auxiliary data that do not couple to the gravitational wave data, in order to avoid possible false detections due to non-Gaussian noise.
A simple but complete analysis system using {\tt TFCLUSTERS} and satisfying these requirements was developed in \cite{thesis} in order to compute upper limits on the rate of occurrence of events that are expected to radiate gravitational radiation.
A comparison of event lists from the individual detectors was used to carry out this coincidence analysis.
This approach may, however, not give the best network detection efficiency, in part because of the relatively coarse time resolution of {\tt TFCLUSTERS}.
Enhanced versions of the algorithm, designed to analyze coherently data from many detectors, might prove to be more efficient.

\acknowledgments The author wishes to thank Rainer Weiss for helpful discussions and comments on the work presented here, and the LSC group at the University of Wisconsin-Milwaukee for generous time allocation on their computer system. This work was supported by the National Science Foundation under cooperative agreement PHY-9210038 and the award PHY-9801158, and by the {\it Fonds pour la Formation de Chercheurs et l'Aide \`a la Recherche} of the Province of Qu\'ebec, Canada.

\appendix
\section{Proof of theorem 1} \label{proof_th1}
Consider first the deterministic equivalent to Eq. (\ref{eq:noise_model_tdom}):
\begin{equation}
\vc{y} = \vc{s} + \delta\vc{u},\label{eq:det_power_model}
\end{equation}
where $\vc{u}$ is a nuisance parameter, so that the equivalent to Eq. (\ref{eq:power_gauss}) is
\begin{equation}
P_{ij}(\vc{y}) = |\tilde{s}_{ij} + \delta\tilde{u}_{ij}|^2, \label{eq:det_power_model_tfdom}
\end{equation}
where $\tilde{u}_{ij} \in \Complex$ and where by definition $|\tilde{u}_{ij}| \leq 1$.

\noindent
{\it Lemma 1:}\\
For the model described by Eq. (\ref{eq:det_power_model}) with $\delta^2 = \eta$, $\forall \vc{s} \in \Real^N$ and $\forall \vc{y}$ satisfying Eq. (\ref{eq:det_power_model}), $\hat{Q}(\vc{y}) \leq Q$.\\
{\it Proof:}\\
Consider Eq. (\ref{eq:power_threshold}).
Trivially, if $|\hat{\tilde{s}}_{ij}|^2 = 0$, then $|\hat{\tilde{s}}_{ij}|^2 \leq |\tilde{s}_{ij}|^2$.
For $|\hat{\tilde{s}}_{ij}|^2 > 0$, $|\hat{\tilde{s}}_{ij}|^2 = P_{ij}(\vc{y}) - \eta$. 
From Eq. (\ref{eq:det_power_model_tfdom}), $|\tilde{s}_{ij}|^2 \geq P_{ij}(\vc{y}) - \delta^2$.
Since $\delta^2 = \eta$, this gives $|\tilde{s}_{ij}|^2 \geq |\hat{\tilde{s}}_{ij}|^2$ for all $\vc{s}, \vc{y}$.
Summing over $i$ and $j$ gives $\hat{Q} \leq Q$.\\

\noindent
{\it Lemma 2:}\\
Given Eq. (\ref{eq:det_power_model}) with $\delta^2 = \eta$, $\forall \vc{s} \in \Real^N$, and $\forall \vc{y}$ respecting Eq. (\ref{eq:det_power_model}), $\hat{Q}(\vc{y}) \geq \hat{q}(\vc{y})$, where $\hat{q}(\vc{y})$ is any power estimator satisfying 
\begin{equation}
\hat{q}(\vc{y}') \leq Q, \mbox{$\forall\vc{s}' \in \Omega(\vc{s})$ and $\forall \vc{y}'$ respecting Eq. (\ref{eq:det_power_model})}, \label{eq:powerCond}
\end{equation}
where 
\begin{equation}
\Omega(\vc{s}) = \left\{\vc{s}' \in \Real^N : |\vc{s} - \vc{s}'|^2 \leq \sum_{ij} \min\left(4\delta^2, 2\delta |\tilde{s}_{ij}| \left[1 + \frac{|\tilde{s}_{ij}|}{\delta} \left( 1 + \sqrt{1 + \frac{2\delta}{|\tilde{s}_{ij}|}} \right) \right] \right) \right\}. \label{eq:Omega}
\end{equation}
{\it Proof:}\\
Suppose $\exists \vc{y}^0$ such that $\hat{q}(\vc{y}^0) > \hat{Q}(\vc{y}^0)$, so $\tilde{y}^0_{ij} = \tilde{s}^0_{ij} + \delta\tilde{u}^0_{ij}$.
One can construct a signal $\vc{s}'$ such that 
\begin{equation}
\tilde{s}'_{ij} =  \left\{ 
\begin{array}{ll}
\tilde{y}^0_{ij} - \delta \tilde{u}'_{ij} & \mbox{if $|\tilde{y}^0_{ij}| > \delta$} \\
 0 & \mbox{otherwise}.
\end{array}
\right.
\end{equation}
The freedom provided by $\tilde{u}'_{ij} \in \Complex$ is used to choose all values of $\tilde{u}'_{ij}$ so that $\tilde{u}'_{ij}$ and $\tilde{s}'_{ij}$ are orthogonal.
If $\tilde{s}'_{ij} = 0$, the direction of $\tilde{u}'_{ij}$ is unimportant.
For all $i,j$, $|\tilde{u}'_{ij}| = 1$.
Note that this choice of $\tilde{u}'_{ij}$ and $\tilde{s}'_{ij}$ satisfies Eq. (\ref{eq:det_power_model}).

Also, since $\tilde{s}'_{ij} - \tilde{s}^0_{ij} = \delta(\tilde{u}^0_{ij} - \tilde{u}'_{ij})$, $|\tilde{s}'_{ij} - \tilde{s}^0_{ij}|^2 \leq 4\delta^2$.
Now, $|\tilde{y}^0_{ij}|^2 = \delta^2 + |\tilde{s}'_{ij}|^2$, so,
\begin{eqnarray}
|\tilde{s}'_{ij} - \tilde{s}^0_{ij}|^2 = |\tilde{s}^0_{ij}|^2 + |\tilde{s}'_{ij}|^2 - 2 \langle \tilde{s}^0_{ij}, \tilde{s}'_{ij}\rangle \\
\leq 2|\tilde{s}^0_{ij}|^2 + 2|\tilde{s}^0_{ij}|\delta + 2|\tilde{s}^0_{ij}|^2 \sqrt{1 + \frac{2\delta}{|\tilde{s}^0_{ij}|}}.
\end{eqnarray}
Combining these inequalities and summing over $i,j$ shows that $\vc{s}' \in \Omega(\vc{s})$.

Then,
\begin{eqnarray}
\hat{Q}(\vc{y}^0) = \sum_{i,j}(P_{ij}(\vc{y}^0) - \eta)_+ \\
& = \sum_{i,j}(|\tilde{s}'_{ij} + \delta \tilde{u}'_{ij}|^2  - \eta)_+ \\
& = \sum_{i,j}(|\tilde{s}'_{ij}|^2 + \delta^2 - \eta) \\
& = |\vc{s}'|^2.
\end{eqnarray}

It follows that $\hat{q}(\vc{y}^0) > \hat{Q}(\vc{y}^0)$ implies $\hat{q}(\vc{y}^0) > |\vc{s}'|^2$, contradicting Eq. (\ref{eq:powerCond}).\\

\noindent
{\it Lemma 3:}\\
For the model described by Eq. (\ref{eq:det_power_model}) with $\delta^2 = \eta$, $\forall \vc{s} \in \Real^N$ and $\forall \vc{y}$ satisfying Eq. (\ref{eq:det_power_model}), $\hat{Q}(\vc{y}) \geq Q - \sum_{i,j}\min(2\delta^2, |\tilde{s}_{ij}|^2)$.\\
{\it Proof:}\\
Obviously, $|\tilde{s}_{ij}|^2 - |\hat{\tilde{s}}_{ij}|^2 \leq |\tilde{s}_{ij}|^2$.
Also, from Eq. (\ref{eq:det_power_model_tfdom}), $P_{ij}(\vc{y}) \geq |\tilde{s}_{ij}|^2 - \delta^2$, so from Eq. (\ref{eq:Pthresh}), $|\hat{\tilde{s}}_{ij}|^2 = P_{ij}(\vc{y}) - \delta^2 \geq |\tilde{s}_{ij}|^2 - 2\delta^2$ when $|\hat{\tilde{s}}_{ij}|^2 > 0$.
When $|\hat{\tilde{s}}_{ij}|^2 = 0$, $\delta^2 > P_{ij}(\vc{y}) \geq |\tilde{s}_{ij}|^2 - \delta^2$, so $|\tilde{s}_{ij}|^2 < 2\delta^2$.
Combining the three inequalities gives $|\tilde{s}_{ij}|^2 - |\hat{\tilde{s}}_{ij}|^2 \leq \min(2\delta^2, |\tilde{s}_{ij}|^2)$; summing over $i,j$ proves the lemma.\\

\noindent
{\it Proof of theorem 1:}\\
In the noise model from Eq. (\ref{eq:power_gauss}), the power in the noise is distributed exponentially:
\begin{equation}
p_{|\tilde{n}_{ij}|^2}(P) = \frac{e^{{-P}/\sigma^2}}{\epsilon^2}.
\end{equation}
The corresponding cumulative probability distribution is
\begin{equation}
P_{|\tilde{n}_{ij}|^2}(P) = 1 - e^{{-P}/\epsilon^2}.
\end{equation}
Hence, the probability that the maximum of $N$ independent realizations of that random variable is less than a threshold $C_N$ is
\begin{equation}
\pi_N \equiv Pr\left(\max_{i,j}|\tilde{n}_{ij}|^2 < C_N\right) = \left( 1 - e^{{-C_N}/\epsilon^2} \right)^N.
\end{equation}
For $N \gg 1$, this can be rewritten as
\begin{equation}
\pi_N = \exp\left(-e^{\log N - C_N/\epsilon^2}\right)
\end{equation}
for $C_N / \epsilon^2 > \log N$.
Hence, for $C_N \rightarrow \epsilon^2\log N$, this probability goes to $1/e$ as $N$ becomes large.
Moreover, any values of $\pi_N > 1/e$ can be achieved with the right choice of $C_N$; for $C_N = \beta\epsilon^2\log N$, $\beta > 1$, $\pi_N \rightarrow 1$ for $N \gg 1$.

It should be noted that the event 
\begin{equation}
\max_{i,j}|\tilde{n}_{ij}|^2 < \beta\epsilon^2\log N
\end{equation}
implies a realization of the noise which can be mimicked by the deterministic noise model [Eq. (\ref{eq:det_power_model})] with $\delta = \epsilon \sqrt{\beta\log N}$.
Hence, the three statements of theorem 1 follow from the three lemmata proved in this appendix.\\

\section{The Narrow-Band Signal Example} \label{narrow_band}
Given a $u$ by $v$ matrix $\vc{Q}$ with elements representing the distribution of the power in the signal in a rectangular sub-region of the time-frequency plane, 
the signal-to-noise ratio is defined by:
\begin{equation}
\rho^2 = \frac{\sum_{i,j} Q_{ij}}{uv\epsilon^2}.
\end{equation}
The elements of the matrix $\vc{D}$ representing the black pixel probability corresponding to the matrix $\vc{Q}$ are given by the integral of the density from Eq. (\ref{eq:rice_prob_signal}):
\begin{equation}
D_{ij} = \int_{\eta}^\infty p_{P_{ij}}(P|Q_{ij}) dP.
\end{equation}

In general, not all pixels where some signal power is present will be black, and the signal will be detected only when a number of black pixels equal or larger than the threshold $\sigma$ form a connected cluster, or when a pair of smaller clusters are close enough.
The contribution to the probability of detection of the signal of such a configuration will be the product of the black pixel probabilities ($D_{ij}$) and of the white pixel probabilities ($1-D_{ij}$) for ``holes''.
Although noise fluctuations could in principle help the detection of a signal by forming a ``bridge'' over regions where no signal is present, summing over these possibilities involve $2^{uv}$ terms.
A slight underestimate of the probability of detection is instead used by summing only over the $n$ pixels where signal is present ($n \leq uv$), which reduces the number of terms to be considered to 
\begin{equation}
\sum_{n_H = 0}^{n - \sigma} \frac{n!}{(n-n_H)!n_H!},
\end{equation}
where $n_H$ is the number of holes.
Of course, the enumeration process can be greatly simplified when the distribution of the power in the signal has some specific symmetries.

\begin{figure}
\begin{center}
\ieps{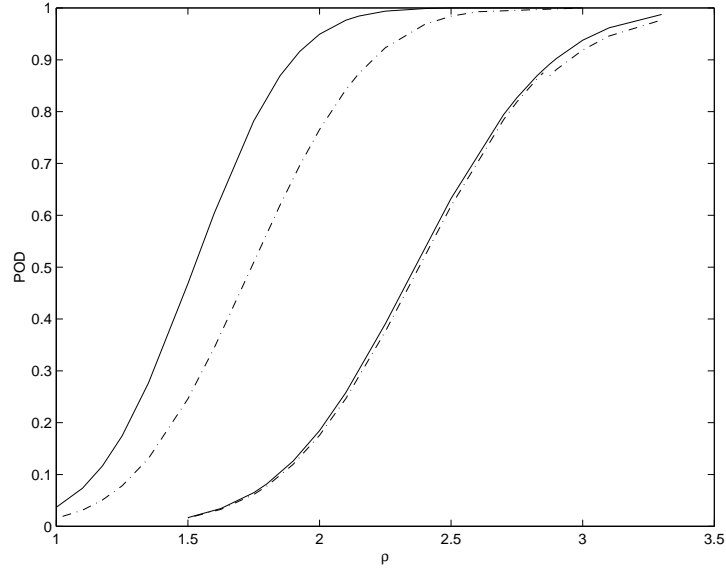}
\caption{The optimized probability of detection of the $6T$ long signal with bandwidth $1/T$ as a function of its signal to noise ratio $\rho$, for a false alarm rate $\lambda = 1/3600T$, and a search bandwidth $512/T$. 
The two rightmost curves correspond to the {\tt TFCLUSTERS} detector.
For $\rho < 2.875$, the optimal thresholds are: $\sigma = 3$, 
$\vc{\delta} = [0,0,3]$. 
For $\rho>2.875$, they are: $\sigma = 4$, $\vc{\delta} = [0,0,0,0,2,0]$.
The two leftmost curves correspond to the ideal power detector.
The solid lines are for the case where the arrival matches exactly the spectrogram gridding (i.e., the signal occupies six pixels), while the dash-dotted lines correspond to the POD for the general case (power spread over seven pixels).
}
\label{fig:SNR_POD}
\end{center}
\end{figure}

Consider now the example of section \ref{clustering_POD}.
Under the assumption that the starting time of the signal matches exactly the binning of the spectrogram used to detect it, it will be represented by a row of 6 pixels, each with an equal probability $p$ to be black, neglecting effects such as power leakage.
The columns labeled ``POD ($r\rightarrow p, s\rightarrow 0$)'' in tables \ref{tab:ex_qq} and \ref{tab:ex_qq2} give the contributions to the probability of detection (POD) of this signal from various thresholds.
The problem is slightly more complex when the arrival time is taken to be random.
In that case, the signal is spread in general over 7 pixels, with the central five having a black pixel probability $p$, and the leftmost and rightmost having smaller probabilities $r$ and $s$, respectively.
$r$ and $s$ are given by:
\begin{equation}
r = \int_{\eta}^\infty {p_{P_{ij}}(P|P'_s) dP}
\end{equation}
and
\begin{equation}
s = \int_{\eta}^\infty {p_{P_{ij}}(P|P_s - P_s') dP},
\end{equation}
where $P_s$ is the power in the central five pixels, $P'_s$ is the power in the leftmost pixel, and $p_{P_{ij}}(P|Q)$ is given by Eq. (\ref{eq:rice_prob_signal}).
The POD is then given by
\begin{equation}
\frac{1}{P_s} \int_{0}^{P_s} {{\rm POD}(r(P'_s), s(P_s-P'_s)) dP'_s},
\end{equation}
where ${\rm POD}(r(P'_s), s(P_s-P'_s))$ is the probability of detection for the 7 pixels configuration; contributions to it from various thresholds can be found in the columns labeled ``POD'' in  tables \ref{tab:ex_qq} and \ref{tab:ex_qq2}.
For the ideal power detector with $\lambda = 1/3600T$, $\tau = 6T$ and $B/\phi=512$, Eq. (\ref{eq:false_alarm_power}) gives $\kappa \approx 23.87\epsilon^2$ for $V = uv = 6$ and $\kappa \approx 25.55\epsilon^2$ for $V = uv = 7$.
Eq. (\ref{eq:detection_prob_power}) is then used directly to generate the ideal power detector curves of figure \ref{fig:SNR_POD}.

%%%%%%%%%%%%%%%%%%%%%%%TABLES%%%%%%%%%%%%%%%%%%%%%%%%%%%%%%%%%
\begin{table}
\begin{center}
\begin{tabular}{|c|c|c|c|c|c|c|c|c|c|c|c|c|c|c|c|c|c|} \hline
\multicolumn{3}{|c|}{} & \multicolumn{15}{c|}{$R$} \\ \hline
$S_1$	&	$S_2$	&	$d$	& 6 & 7 & 8 & 9 & 10 & 11 & 12 & 13 & 14 & 15 & 16 & 17 & 18 & 19 & 20 \\ \hline

1 & 1 & 2 & 4 & 4 & & & & & & & & & & & & & \\ \hline
  &   & 3 &   &   & 12 & & & & & & & & & & & & \\ \hline
  &   & 4 &   &   & 16 & & & & & & & & & & & & \\ \hline \hline

1 & 2 & 2 & & & 24 & 36 & & & & & & & & & & & \\ \hline 
  &   & 3 & & &    &    & 84 & & & & & & & & & & \\ \hline 
  &   & 4 & & &    &    & 108 & & & & & & & & & & \\ \hline \hline

1 & 3 & 2 & & & & 80 & 128 & 64 & & & & & & & & & \\ \hline 
  &   & 3 & & & &    &     & 240 & 128 & & & & & & & & \\ \hline 
  &   & 4 & & & &    &     & 304 & 160 & & & & & & & & \\ \hline 
  &   & 5 & & & &    &     & 368 & 192 & & & & & & & & \\ \hline 
  &   & 6 & & & &    &     & 432 & 224 & & & & & & & & \\ \hline \hline

1 & 4 & 2 & & & &    & 260 & 480 & 360 & 100 & & & & & & & \\ \hline
  &   & 3 & & & &    &     &     & 720 & 680 & 180 & & & & & & \\ \hline
  &   & 4 & & & &    &     &     & 900 & 840 & 220 & & & & & & \\ \hline
  &   & 5 & & & &    &     &     &1080 &1000 & 260 & & & & & & \\ \hline
  &   & 6 & & & &    &     &     &1260 &1160 & 300 & & & & & & \\ \hline \hline

2 & 2 & 2 & & & & & 40 & 56 & & & & & & & & & \\ \hline
  &   & 3 & & & & &     &    & 128 & & & & & & & & \\ \hline
  &   & 4 &  & & & &    &    & 160 & & & & & & & & \\ \hline \hline

2 & 3 & 2 & & & & &     & 240 & 400 & 160 & & & & & & & \\ \hline
  &   & 3 & & & & &     &    &     & 680 & 360 & & & & & & \\ \hline
  &   & 4 & & & & &     &    &     & 840 & 440 & & & & & & \\ \hline 
  &   & 5 & & & & &     &    &     &1000 & 520 & & & & & & \\ \hline 
  &   & 6 & & & & &     &    &     &1160 & 600 & & & & & & \\ \hline \hline

2 & 4 & 2 & & & & &     &    & 792 &1344 & 984 & 216 & & & & & \\ \hline 
  &   & 3 & & & & &     &    &     &     &1944 &1824 & 480 & & & & \\ \hline
  &   & 4 & & & & &     &    &     &     &2376 &2208 & 576 & & & & \\ \hline 
  &   & 5 & & & & &     &    &     &     &2808 &2592 & 672 & & & & \\ \hline
  &   & 6 & & & & &     &    &     &     &3240 &2976 & 768 & & & & \\ \hline \hline 

3 & 3 & 2 & & & & &     & 12 & 324 & 684 & 456 & 108 & & & & & \\ \hline 
  &   & 3 & & & & &     &    &     &     & 864 & 912 & 240 & & & & \\ \hline 
  &   & 4 & & & & &     &    &     &     &1056 &1104 & 288 & & & & \\ \hline 
  &   & 5 & & & & &     &    &     &     &1248 &1296 & 336 & & & & \\ \hline 
  &   & 6 & & & & &     &    &     &     &1440 &1488 & 384 & & & & \\ \hline \hline 

3 & 4 & 2 & & & & &     &    & 112 &2016 &4424 &4172 &1736 & 280 &   & & \\ \hline
  &   & 3 & & & & &     &    &     &     &     &4788 &7000 &3528 & 616 & & \\ \hline 
  &   & 4 & & & & &     &    &     &     &     &5796 &8400 &4200 & 728 & & \\ \hline 
  &   & 5 &  & & & &    &    &     &     &     &6804 &9800 &4872 & 840 & & \\ \hline 
  &   & 6 & & & & &     &    &     &     &     &7812 &11200&5544 & 952 & & \\ \hline \hline

4 & 4 & 2 & & & & &     &    &     & 272 &2912 &7216 &8112 &4800 &1440 & 176 & \\ \hline
  &   & 3 & & & & &     &    &     &     &     &     &6480 &12096&8800 &2944 & 384 \\ \hline
  &   & 4 & & & & &     &    &     &     &     &     &7776 &14400&10400&3456 & 448 \\ \hline
  &   & 5 & & & & &     &    &     &     &     &     &9072 &16704&12000&3968 & 512 \\ \hline
  &   & 6 & & & & &     &    &     &     &     &     &10368&19008&13600&4480 & 576 \\ \hline
\end{tabular}
\caption{Coefficients $k_{S_1 S_2 R}(d)$, multiplied by $S_1+S_2$, for clusters of size $S_1$ and $S_2$, with total perimeter $R$, and separated by a distance $d$.} 
\label{tab:correl_coeff}
\end{center}
\end{table}

\begin{table}
\begin{center}
\begin{tabular}{|c|c|c|} \hline
$S_1$ & $S_2$ & $\langle \nu_{S_1,S_2} (d) \rangle$, $d>2$ \\ \hline
1 & 1 & $2d p^2 q^8$ \\ \hline
1 & 2 & $4 (1+2d) p^3 q^{10}$\\ \hline
1 & 3 & $4 (3+4d) p^4 q^{11} + 8(1+d) p^4 q^{12}$\\ \hline
1 & 4 & $36 (1+d) p^5 q^{12} + 8(5+4d) p^5 q^{13} + 4(3+2d)p^5 q^{14}$ \\ \hline
2 & 2 & $8 (d+1) p^4 q^{12}$ \\ \hline
2 & 3 & $8 (5+4d) p^5 q^{13} + 8(3+2d) p^5 q^{14}$ \\ \hline
2 & 4 & $36(3+2d) p^6 q^{14} + 16(7+4d) p^6 q^{15} + 16(2+d)p^6 q^{16}$ \\ \hline
3 & 3 & $16 (3+2d) p^6 q^{14} + 8(7+4d)p^6 q^{15} + 8(2+d)p^6 q^{16}$ \\ \hline
3 & 4 & $36 (7+4d) p^7 q^{15} + 200(2+d)p^7 q^{16} + 24 (9+4d) p^7 q^{17} + 8(5+2d)p^7 q^{18} $ \\ \hline
4 & 4 & $162(2+d) p^8 q^{16} + 72 (9+4d) p^8 q^{17} + 100 (5+2d) p^8 q^{18} + 16 (11+4d) p^8 q^{19} + 8(3+d) p^8 q^{20}$  \\ \hline 
\end{tabular}
\caption{General formulae for $\langle \nu_{S_1,S_2} (d) \rangle$ for distances $d>2$, for different cluster sizes $S_1$ and $S_2$, as deduced from table \ref{tab:correl_coeff}. By definition, $q \equiv 1 -p$.} 
\label{tab:twopts}
\end{center}
\end{table}

\begin{table}
\begin{center}
\begin{tabular}{|c|c|c|} \hline
$\sigma$	&	POD ($r\rightarrow p, s\rightarrow 0$) & POD 	\\ \hline
7	&		&	$p^5rs$ \\ \hline
6	&	$p^6$ 	&	$p^5(r+s-2rs)$ \\ \hline
5	&	$2p^5q$ &	$p^4(p(1-r)(1-s)+q(r+s))$	\\ \hline
4	&	$p^4q(2+q)$ &	$p^3q(r+s-p(r+s-2))$ \\ \hline
3	&	$p^3q(2+2q)$	& $p^2q(r+s-p(r+s-q+prs-2))$ \\ \hline
2	&	$p^2q(2-2p^3+3q-3p^2q)$	&	
$pq(r+p^3(r(s-1)-s-1)+2p^2q(r(s-1)-s)+s-p(r+s-2+q(qrs-2)))$ \\ \hline
\end{tabular}
\caption{Contributions to the probability of detection (POD) for different thresholds on the cluster size, for a straight line of 6 pixels, all having equal black pixel probability $p$ (columns labeled ``POD ($r\rightarrow p, s\rightarrow 0$)''), and for a row of 7 pixels, with 5 central pixels having probability $p$, and the leftmost and rightmost pixels having probabilities $r$ and $s$, respectively (columns labeled ``POD''). By definition, $q \equiv 1 -p$.}
\label{tab:ex_qq}
\end{center}
\end{table}

\begin{table}
\begin{center}
\begin{tabular}{|c|c|c|c|c|} \hline
$S_1$ & $S_2$ &	$\delta_{S_1,S_2}$ &	POD ($r\rightarrow p, s\rightarrow 0$) & POD \\ \hline
3 & 3 & 2 & & $p^4qrs$ \\ \hline
2 & 4 & 2 & & $2p^4qrs$ \\ \hline
2 & 3 & 2 & $2p^5q$ & $2p^4q(r+s-2rs)$ \\ \hline
  &   & 3 & & $2p^3q^2rs$ \\ \hline
2 & 2 & 2 & $2p^4q^2$ & $p^3q(p(r-1)(s-1)+q(r+s))$ \\ \hline
  &   & 3 & $p^4q^2$ & $p^3q^2(r+s-2rs)$ \\ \hline
  &   & 4 & & $p^3q^2rs$ \\ \hline
1 & 5 & 2 & & $2p^4qrs$ \\ \hline
1 & 4 & 2 & $2p^5q$ & $p^4q(r(2-3s)+2s)$\\ \hline
  &   & 3 & & $2p^3q^2rs$ \\ \hline
1 & 3 & 2 & $4p^4q^2$ & $p^3q(2p(1-r)(1-s)+q(r(2-s)+2s))$\\ \hline
  &   & 3 & $2p^4q^2$ & $2p^3q^2(r+s-2rs)$\\ \hline
  &   & 4 & $2p^2q^3rs$ & \\ \hline
1 & 2 & 2 & $3p^3q^2(1+q)$& $p^2q^2(q(2r(1-s)+3s)+p(4-r-s-2rs))$ \\ \hline
  &   & 3 & $4p^3q^3$ & $p^2q^2(2p(1-r)(1-s)+q(2r+s))$ \\ \hline
  &   & 4 & $2p^3q^3$ & $2p^2q^3(r+s-2rs)$ \\ \hline
  &   & 5 & $2pq^4rs$ & \\ \hline
1 & 1 & 2 & $4p^2q^3$ & $pq^2(p^2(1-r)(1-s)+q^2(r+s)+pq(3-rs))$ \\ \hline
  &   & 3 & $3p^2q^4$ & $pq^3(2p(1-r)(1-s)+q(r+s-rs))$ \\ \hline
  &   & 4 & $2p^2q^4$ & $pq^3(p(1-r)(1-s)+q(r+s-2rs))$\\ \hline
  &   & 5 & $p^2q^4$  & $p q^4(r+s-2rs)$\\ \hline
  &   & 6 & & $q^5rs$ \\ \hline
\end{tabular}

\caption{Same as table \ref{tab:ex_qq} for thresholds on the distance between two clusters of size $S_1$ and $S_2$.}
\label{tab:ex_qq2}
\end{center}
\end{table}

%%%%%%%%%%%%%%%%%%%%%%FIGURES%%%%%%%%%%%%%%%%%%%%%%%%%%%

\begin{references}
\bibitem{whitening} 
An arbitrary random process can always be reduced to a set of uncorrelated coefficients (``whitened'') by the Karhunen-Lo\`eve expansion.
Signals are affected by the whitening; however, the only place where this could have an adverse effect here is if the whitening were reducing the amount of clustering of astrophysical signals.
\bibitem{Fisher_estimation} The Neyman-Pearson lemma shows that Eq. (\ref{eq:match_filter}) is optimal when choosing between Eqs. (\ref{eq:noise_model_tdom}) and (\ref{eq:just_noise}); for the model $y(t) = A s(t) + \epsilon n(t)$ with $|s(t)|=1$, Eq. (\ref{eq:match_filter}) is the most efficient estimator for $A$, in the sense of being unbiased and of achieving the minimum possible mean square error (from the Fisher information, this is $\epsilon^2$): $\hat{A} = A - \epsilon^2 \partial \log p[y(t)|A] / \partial A = \langle y(t), s(t) \rangle$.
\bibitem{Ito} Rules to use the It\^o integral $\Itoint_0^T b(t)w(t)dt = \Itoint_0^T b(t)dw(t)$ are similar to those for normal integration, except that second order terms in the increment $dw(t)$ are preserved, corresponding to the fact the these increments are $O(dt^{1/2})$ for a Wiener process. See section II of \cite{kailath} for a pedagogical introduction.
\end{references}
\end{document}